\documentclass[aps,prb,twocolumn,groupedaddress,showpacs]{revtex4}

\usepackage[dvips]{graphicx}
\usepackage{epsfig}

\begin{document}

\newcommand\qu{{\bf q}}
\newcommand\nalpha{{\{n_\alpha\}}}
\newcommand\nqu{{\{n_\qu\}}}
\newcommand\nalphap{{\{n_{\alpha'}\}}}
\newcommand\nalphas{{\{n_{\alpha''}\}}}
\newcommand\nqup{{\{n_{\qu'}\}}}
\newcommand\nalphaone{{\{n_{\alpha_1}\}}}
\newcommand\nalphatwo{{\{n_{\alpha_2}\}}}
\newcommand\nalphaonep{{\{n_{\alpha'_1}\}}}
\newcommand\nalphatwop{{\{n_{\alpha'_2}\}}}
\newcommand\rmtr{{\rm tr}}

%\preprint{Submitted to Phys. Rev. B}

\title{
Quantum-transport theory for semiconductor nanostructures: \\
A density-matrix formulation }


\author{Rita C. Iotti}
%\email[]{Rita.Iotti@INFM.PoliTo.It}
%\thanks{}
\author{Emanuele Ciancio}
%\email[]{Emanuele.Ciancio@PoliTo.It}
%\thanks{}
\author{Fausto Rossi}
\email[]{Fausto.Rossi@PoliTo.It}
%\thanks{}
\affiliation{
Istituto Nazionale per la Fisica della Materia (INFM) and
Dipartimento di Fisica, Politecnico di Torino,
Corso Duca degli Abruzzi 24,
10129 Torino, Italy
}


\date{\today}

\begin{abstract}

A general density-matrix formulation of quantum-transport phenomena in semiconductor
nanostructures is presented. More specifically, contrary to the conventional single-particle correlation
expansion, we shall investigate separately the effects of the adiabatic or Markov limit and of the reduction
procedure. Our fully operatorial approach allows us to better identify the general properties of the scattering
superoperators entering our effective quantum-transport theory at various description levels, e.g., $N$
electrons-plus-quasiparticles, $N$ electrons only, and single-particle picture.
In addition to coherent transport phenomena characterizing the transient response of the system, the proposed
theoretical description allows to study scattering induced phase coherence in steady-state conditions.
As prototypical example, we shall investigate polaronic effects in strongly biased semiconductor superlattices.

\end{abstract}

%\keywords{}

\maketitle

%\section{}
%\subsection{}
%\subsubsection{}

% If in two-column mode, this environment will change to single-column
% format so that long equations can be displayed. Use
%\begin{widetext}
%\end{widetext}

\section{Introduction}\label{s-intro}

Recent developments in nanoscience/technology
pushes device miniaturization toward limits where the traditional
semiclassical transport treatments~\cite{MC} can no longer be
employed, and more rigorous quantum-transport approaches are
imperative.\cite{QT}
However, in spite of the
quantum-mechanical nature of carrier dynamics in the core region
of typical nanostructured devices
---like semiconductor superlattices, double-barrier structures,
and quantum dots--- the overall behavior of such quantum systems
is often the result of a non-trivial in\-ter\-play between phase
coherence and energy relaxation/dephasing.\cite{RMP}  It follows
that a proper treatment of such novel nanoscale devices requires a
theoretical modelling able to properly account for both coherent
and incoherent ---i.e., phase-breaking--- processes on the same
footing within a many-body picture.

More precisely, the idealized behavior of a so-called ``quantum
device''~\cite{Cbook} is usually described via the elementary
physical picture of the square-well potential and/or in terms of a
simple quantum-mechanical $n$-level system. For a quantitative
investigation of state-of-the-art quantum optoelectronic devices,
however, two features strongly influence and modify such
simplified scenario: (i) the intrinsic many-body nature of the
carrier system under investigation, and (ii) the potential
coupling of the electronic subsystem of interest with a variety of
interaction mechanisms, including the presence of spatial
boundaries.\cite{WF,PRL-QT,BR}

The wide family of so-called quantum devices can be divided into
two main classes: a first one grouping semiconductor devices
characterized by a genuine quantum-mechanical behavior of their
carrier subsystem, and a second one which comprises
low-dimensional nanostructures whose transport dynamics may be
safely treated within the semiclassical picture.

Devices within the first class ---characterized by a weak coupling
of the carrier subsystem with the host material--- are natural
candidates for the implementation of quantum
information/computation processing.\cite{QIC}  These include, in
particular, semiconductor quantum-dot structures,\cite{QDs} for
which all-optical implementations have been recently
proposed.\cite{ZR,Biolatti,IEEE}  In this case, the pure
quantum-mechanical carrier dynamics is only weakly disturbed by
decoherence processes; therefore, the latter are usually described
in terms of extremely simplified models.

Conversely, quantum devices in the second class ---in spite of
their partially discrete energy spectrum due to spatial carrier
confinement--- exhibit a carrier dynamics which can be still
described via a semiclassical scattering picture. Such
optoelectronic nanostructured devices include multi-quantum-well
and superlattice structures, like quantum-cascade lasers
(QCLs).\cite{QCL1,QCLr} These systems are characterized by a
strong interplay between coherent dynamics and
energy-relaxation/dephasing processes; it follows that for a
quantitative description of such non-trivial coherence/dissipation
coupling the latter need to be treated via fully microscopic
models.

In this paper we shall primarily focus on this second class of
quantum devices, providing a comprehensive microscopic theory of
charge transport in semiconductor nanostructures based on the
well-known density-matrix approach. It is worth mentioning that an
alternative approach, equivalent to the density-matrix formalism
employed in this paper, is given by the nonequilibrium Green's
function technique; the latter can be regarded as an extension of
the well-known equilibrium or zero-temperature Green's function
theory to nonequilibrium regimes, introduced in the 1960s by
Kadanoff and Baym~\cite{KB} and Keldysh.\cite{K}  An introduction
to the theory of nonequilibrium Green's functions with
applications to many problems in transport and optics of
semiconductors can be found in the book by Haug and
Jauho.\cite{HJ} By employing ---and further developing and
extending--- such nonequilibrium Green's function formalism, a
number of groups have proposed efficient quantum-transport
treatments for the study of various semiconductor
nanostructures~\cite{Wacker-rev} as well as of modern
micro/optoelectronic devices.~\cite{nemo}

Within the general density-matrix formalism two
different strategies are commonly employed: (i) the
quantum-kinetic treatment,\cite{RMP} and (ii) the description
based on the Liouville-von Neumann equation.\cite{RBJ}

The primary goal of a quantum-kinetic theory is to evaluate the
temporal evolution of a reduced set of single- or few-particle
quantities directly related to the electro-optical phenomenon
under investigation, the so-called kinetic variables of the
system. However, due to the many-body nature of the problem, an
exact solution in general is not possible; it follows that for a
detailed understanding realistic semiconductor models have to be
considered, which then can only be treated approximately. Within
the kinetic-theory approach one starts directly with the equations
of motion for the single-particle density matrix. Due to the
many-body nature of the problem, the resulting set of equations of
motion is not closed; instead, it constitutes the starting point
of an infinite hierarchy of higher-order density matrices. Besides
differences related to the quantum statistics of the
quasiparticles involved, this is equivalent to the BBGKY hierarchy
in classical gas dynamics.\cite{BBJKY}  The central
approximation in this formalism is the truncation of the
hierarchy. This can be based on different physical pictures. A
common approach is to use the argument that correlations involving
an increasing number of particles will become less and less
important.\cite{RMP}  An alternative quantum-kinetic
scheme
---based on an expansion in powers of the exciting laser field---
has been introduced by Axt and Stahl, the so-called ``dynamics
controlled truncation'' (DCT).\cite{AM}

Within the treatment based on the Liouville-von Neumann equation,
the starting point is the equation of motion for the global
density-matrix operator, describing many electron plus various
quasiparticle excitations. The physical quantities of interest for
the electronic subsystem are then typically derived via a suitable
``reduction procedure'', aimed at tracing out non-relevant degrees
of freedom. Contrary to the kinetic theory, this approach has
allowed for a fully quantum-mechanical treatment of high-field
transport in semiconductors,\cite{QMC} thus overcoming some of the
basic limitations of conventional kinetic treatments, e.g., the
completed-collision limit and the Markov approximation.

Primary goal of the present paper is to discuss in very general
terms the physical properties and validity limits of the so-called
``adiabatic'' or Markov approximation. Within the traditional
semiclassical or Boltzmann theory, this approximation is typically
introduced together with the so-called diagonal approximation,
i.e., the neglect of non-diagonal density-matrix elements.
However, as described in Ref.~\onlinecite{Tilmann}, the Markov
limit can be also performed within a fully non-diagonal
density-matrix treatment of the problem; this leads to the
introduction of generalized in- and out-scattering superoperators,
whose general properties and physical interpretation are not
straightforward. In particular, it is imperative to understand if
---and under which conditions--- the adiabatic or Markov
approximation preserves the positive-definite character of our
reduced density matrix; indeed, this distinguished property is
generally lost within the quantum-kinetic approaches previously
mentioned.\cite{RMP,Tilmann} To this end, starting from the
Liouville-von Neumann-equation approach, we shall propose a very
general treatment of the Markov approximation. More specifically,
contrary to the conventional single-particle correlation expansion
of the kinetic theory, we shall investigate separately the effects
of the Markov limit and of the reduction procedure. Our fully
operatorial approach will allow us to better identify the general
properties of the scattering superoperators entering our effective
quantum-transport theory at various description levels, e.g., $N$
electrons-plus-quasiparticles, $N$ electrons only, and
single-particle picture. In addition to coherent transport
phenomena characterizing the transient response of the system, the
proposed theoretical description allows to study scattering
induced phase coherence in steady-state conditions. In particular,
based on the proposed approach we shall consider ---as prototypical example--- polaronic coherence in strongly biased semiconductor superlattices.

The paper is organized as follows. In Sect.
\ref{s-TB} we present the proposed theoretical approach: after
specifying the physical system under investigation
(Sect.~\ref{ss-PS}) and recalling the fundamentals of the
density-matrix formalism (Sect.~\ref{ss-DMF}), we shall introduce
the Markov approximation (Sect.~\ref{ss-AMA}) and derive the
explicit form of the scattering superoperators
(Sect.~\ref{ss-GSS}) as well as their semiclassical counterparts
(Sect.~\ref{ss-SL}); we shall then discuss the so-called reduction
procedure (Sect.~\ref{ss-RD}) and the single-particle description
(Sect.~\ref{ss-SPD}) for carrier-carrier as well as
carrier-quasiparticle interactions. In Sect.~\ref{s-PCE} we shall
address the general problem of scattering-induced phase coherence;
in particular, we shall present a few simulated experiments
concerning polaronic coherence in semiconductor superlattices.
Finally, in Sect.~\ref{s-SC} we shall summarize and draw some
conclusions.

\section{Theoretical background}\label{s-TB}

\subsection{Physical system}\label{ss-PS}

To provide a general formulation of quantum charge transport in
semiconductor nanostructures, let us consider a generic carrier
gas within a semiconductor crystal in the presence of
electromagnetic fields. The corresponding Hamiltonian can be
schematically written as
\begin{equation}\label{H}
{\hat H} = {\hat H}_\circ + {\hat H}'\ .
\end{equation}

The first term,
\begin{equation}\label{H0}
{\hat H}_\circ = {\hat H}_\circ^c + {\hat H}_\circ^{qp} =
\sum_\alpha \epsilon_\alpha \hat c^\dagger_\alpha \hat c^{
}_\alpha + \sum_\qu \epsilon_\qu \hat b^\dagger_\qu \hat b^{ }_\qu
\ .
\end{equation}
is the sum of the free-carrier and free-quasiparticle
Hamiltonians, where the Fermionic operators $\hat
c^\dagger_\alpha$ ($\hat c^{ }_\alpha$) denote creation
(destruction) of a carrier in the single-particle state $\alpha$
(with energy $\epsilon_\alpha$), while the Bosonic operators $\hat
b^\dagger_\qu$ ($\hat b^{ }_\qu$) denote creation (destruction) of
a generic quasiparticle excitation with wavevector $\qu$ and
energy $\epsilon_\qu$, i.e., phonons, photons, plasmons, etc.

The second term, ${\hat H}'$, is the sum of all possible
carrier-carrier as well as carrier-quasiparticle interaction
Hamiltonians, i.e., car\-rier-pho\-non, carrier-photon,
carrier-plasmon, etc.

The noninteracting carrier-plus-quasiparticle ba\-sis sta\-tes are given by the eigenstates of
${\hat H}_\circ$:
the generic eigenstate
$
\vert \lambda \rangle = \vert \nalpha \rangle \otimes \vert \nqu \rangle
$
is the tensor product of noninteracting carrier and quasiparticle states corresponding, respectively,
to the occupation numbers $\nalpha$ and $\nqu$, while the noninteracting energy spectrum
$
\epsilon_\lambda = \sum_\alpha \epsilon_\alpha n_\alpha + \sum_\qu \epsilon_\qu n_\qu
$
is the sum of the total carrier and quasiparticle energies.

The interaction Hamiltonian ${\hat H}'$ cannot in general be
treated exactly. A typical approach consists of regarding it as a
perturbation acting on the noninteracting
carrier-plus-quasiparticle states $\{\vert \lambda \rangle\}$. In
this context, the basic ingredients are the matrix elements of
${\hat H}'$ within our noninteracting basis states: $
H'_{\lambda\lambda'} = \langle \lambda \vert {\hat H}' \vert
\lambda' \rangle $.

\subsection{Density-matrix formalism}\label{ss-DMF}

In view of the huge number of degrees of freedom ($\{\alpha\}, \{\qu\}$) involved in the microscopic
treatment of any solid-state system, a statistical description of the problem is imperative. As we
shall see (in Sect.~\ref{ss-RD}), this will result in a suitable statistical average over ``non-relevant''
degrees of freedom.

Given a physical quantity $A$ ---de\-scri\-bed by the operator
${\hat A}$--- its quantum plus statistical average value is given
by
\begin{equation}\label{average1}
A = \overline{\langle \psi \vert {\hat A} \vert \psi \rangle } =
{\rm tr}\left\{{\hat A} {\hat \rho}\right\} \ ,
\end{equation}
where
\begin{equation}\label{rho}
{\hat \rho} = \overline{\vert \psi \rangle \langle \psi \vert }
\end{equation}
is the so-called density-matrix operator. The latter is defined as
statistical average of the projection operator corresponding to
the generic state vector $\vert \psi \rangle$ of the system,
%The density-matrix operator ${\hat \rho}$ in (\ref{rho})
and can then be regarded as the statistical generalization of the
quantum-mechanical concept of state vector.

Starting from the global Schr\"odinger equation describing our
interacting carrier-plus-quasiparticle many-body system, the
following Liouville-von Neumann equation of motion for the
density-matrix operator can be readily obtained:
\begin{equation}\label{LvN1}
{d{\hat \rho} \over dt} = {\cal L}\left(\hat\rho\right) = {1 \over i\hbar} \left[{\hat H}, {\hat \rho}\right]\ ,
\end{equation}
where ${\cal L}$ is usually referred to as Liouville
superoperator. Equation~(\ref{LvN1}) can be regarded as the
statistical generalization of the Schr\"odinger equation; Its
exact solution is given by
\begin{equation}\label{exact}
{\hat \rho}(t) = e^{{\cal L}(t-t_0)} \hat\rho(t_0) = {\hat U}^{ }(t-t_0) {\hat \rho}(t_0)
{\hat U}^\dagger(t-t_0) \ ,
\end{equation}
where
\begin{equation}\label{U}
{\hat U}(t-t_0) = e^{{\hat H}(t-t_0) \over i \hbar}
\end{equation}
is the evolution operator corresponding to the total Hamiltonian
${\hat H}$ in (\ref{H}). Such exact solution corresponds to a
fully quantum-mechanical unitary evolution of the whole many-body
system, i.e., no energy relaxation/dephasing. Indeed, it is easy
to verify that the total quantum entropy
\begin{equation}\label{entropy}
S = -k_{\rm B}\, \rmtr\left\{{\hat \rho} \log{\hat \rho}\right\}
\end{equation}
is not affected by the unitary transformation in
(\ref{U}).\cite{note-entropy}

As anticipated, the total many-body Hamiltonian in (\ref{H})
cannot be treated exactly. Aim of a quantum-transport theory is to
derive effective equations describing the carrier subsystem of
interest within some approximation scheme; this is typically
realized via the following two basic steps: first an adiabatic
decoupling between the different time-scales induced by ${\hat
H}_\circ$ and ${\hat H}'$ ---called Markov limit--- and then a
projection of the global system dynamics over a subsystem of
interest via the introduction of a so-called reduced
density-matrix operator.

\subsection{The adiabatic or Markov approximation}\label{ss-AMA}

Starting from the separation ${\hat H} = {\hat H}_\circ + {\hat H}'$ in (\ref{H}), the Liouville-von
Neumann equation (\ref{LvN1}) can be written as
\begin{equation}\label{LvN2}
{d{\hat \rho} \over dt} =
{d{\hat \rho} \over dt}\Bigl|_{{\hat H}_\circ} +
{d{\hat \rho} \over dt}\Bigl|_{{\hat H}'} \ ,
\end{equation}
where the two contributions describe, respectively, the time evolution induced by the noninteracting Hamiltonian
${\hat H}_\circ$ and by the interaction term ${\hat H}'$.

The first contribution can be treated exactly within the standard interaction scheme.
Indeed, it is easy to show that the time evolution of the density-matrix operator in the interaction picture,
\begin{equation}\label{rho_i}
{\hat \rho}^i = {\hat U}^\dagger_\circ(t-t_0) {\hat \rho} {\hat U}^{ }_\circ(t-t_0)\ ,
\end{equation}
is simply given by
\begin{equation}\label{LvN_i}
{d{\hat \rho}^i \over dt} = -i \left[\hat{\cal H}^i, {\hat \rho}^i\right]\ ,
\end{equation}
where ${\hat U}^{ }_\circ(t-t_0)$ is the evolution operator
corresponding to the noninteracting Hamiltonian ${\hat H}_\circ$,
and
\begin{equation}
\label{calH_i} \hat{\cal H}^i(t)= {\hat U}^\dagger_\circ(t-t_0)
\frac{{\hat H}'}{\hbar} {\hat U}^{ }_\circ(t-t_0)\
\end{equation}
denotes the Hamiltonian ${\hat H}'$ in units of $\hbar$ within the
interaction picture.

The key idea beyond any perturbation approach is that the effect
of the interaction Hamiltonian ${\hat H}'$ is ``small'' compared
to the free evolution dictated by the noninteracting Hamiltonian
${\hat H}_\circ$. More precisely, the interaction matrix elements
$H'_{\lambda\lambda'}$ are smaller than the typical energy
difference $\epsilon_\lambda-\epsilon_{\lambda'}$.

Following this spirit, by formally integrating Eq.~(\ref{LvN_i}) from $t_0$ to the current time $t$,
we get:
\begin{equation}\label{FI}
{\hat \rho}^i(t) = {\hat \rho}^i(t_0) -i \int_{t_0}^t dt'
\left[\hat{\cal H}^i(t'), {\hat \rho}^i(t')\right]\ .
\end{equation}
By inserting the above formal solution for ${\hat \rho}^i(t)$ on the right-hand side of Eq.~(\ref{LvN_i})
we obtain an integro-differential equation of the form:
\begin{widetext}
\begin{equation}\label{IDE}
{d \over dt} {\hat \rho}^i(t) = -i \left[\hat{\cal H}^i(t), {\hat \rho}^i(t_0)\right]
-
\int_{t_0}^t dt'
\left[\hat{\cal H}^i(t), \left[\hat{\cal H}^i(t'), {\hat \rho}^i(t')\right]\right]\ .
\end{equation}
\end{widetext}
We stress that so far no approximation has been introduced: Equations (\ref{LvN_i}), (\ref{FI}), and
(\ref{IDE}) are all fully equivalent, we have just isolated the first-order contribution from the full
time evolution in Eq.~(\ref{LvN_i}).
It is then clear that, by iteratively substituting Eq.~(\ref{FI}) into itself, the above procedure can
be extended to any perturbation order.
This leads to the well-known Neuman series:
\begin{widetext}
\begin{equation}\label{Neumann}
\hat{\rho}^i(t) = \hat{\rho}^i(t_0) + \sum_{n = 1}^\infty (-i)^n
\int_{t_0}^t dt_1 \int_{t_0}^{t_1} dt_2 \dots \int_{t_0}^{t_{n-1}}
dt_n [\hat{\cal H}^i(t_1), [\hat{\cal H}^i(t_{2}), \dots
[\hat{\cal H}^i(t_n), \hat{\rho}^i(t_0)]\dots ]] \ .
\end{equation}
\end{widetext}
The latter constitutes the starting point of the quantum Monte
Carlo method for the study of charge-transport phenomena in
semiconductors.\cite{QMC}

In order to introduce the so-called adiabatic or Markov approximation, let us now focus on the time
integral in Eq.~(\ref{IDE}). Here, the two quantities to be integrated over $t'$ are the interaction
Hamiltonian $\hat{\cal H}^i$ and the density-matrix operator ${\hat \rho}^i$. In the spirit of the perturbation
approach previously recalled, the time variation of ${\hat \rho}^i$ can be considered adiabatically slow
compared to that of the Hamiltonian $\hat{\cal H}^i$ within the interaction picture;
indeed, the latter will exhibit rapid oscillations due to the noninteracting unitary transformation
${\hat U}_\circ$.
As a result, the density-matrix operator ${\hat \rho}^i$ can be taken out of the time integral and
evaluated at the current time $t$.

Within such adiabatic limit we get the following effective Liouville-von Neumann equation:
\begin{equation}\label{Markov_i}
{d \over dt} {\hat \rho}^i(t) =
-i \left[\hat{\cal H}^i(t), {\hat \rho}^i(t_0)\right]
-\left[\hat{\cal H}^i(t), \left[\hat{\cal K}^i(t), {\hat \rho}^i(t)\right]\right]
\end{equation}
with
\begin{equation}\label{calK_i}
\hat{\cal K}^i(t) = \int_{t_0}^t \hat{\cal H}^i(t') dt' \ .
\end{equation}
The above equation has still the double-commutator structure in (\ref{IDE}) but it is now local in time.

Going back to the original Schr\"odinger picture, we finally get:
\begin{equation}\label{LvN-eff_L}
{d{\hat \rho} \over dt} = \tilde{\cal L}\left({\hat \rho}\right) +
\hat C \ ,
\end{equation}
where
\begin{equation}\label{Ltilde}
\tilde{\cal L}\left({\hat \rho}\right) = {1 \over i\hbar}
\left[{\hat H}_\circ, {\hat \rho}\right] - \left[\hat{\cal H},
\left[\hat{\cal K},{\hat \rho}\right]\right] \equiv {1 \over
i\hbar} \left[{\hat H}_\circ, {\hat \rho}\right] +
\Gamma\left({\hat \rho}\right) \, ,
\end{equation}
with
\begin{equation}\label{Gamma}
\hat{\cal K} = \int_{t_0}^t dt' {\hat U}^{ }_\circ(t-t') \hat{\cal
H} {\hat U}^\dagger_\circ(t-t') \ ,
\end{equation}
is the effective Liouville superoperator within our approximation
scheme, and
\begin{equation}\label{C}
\hat C(t) = -i \left[\hat{\cal H}, {\hat U}_\circ(t-t_0) {\hat
\rho}(t_0) {\hat U}^\dagger_\circ(t-t_0)\right] \ .
\end{equation}

The time-dependent operator $\hat{C}$ in (\ref{C}) describes how
the quantum-correlation effects at the initial time $t_0$
propagate to the current time $t$; indeed, combining
Eqs.~(\ref{exact}) and (\ref{C}), the latter can be rewritten as:
\begin{equation}\label{C-bis}
\hat C(t) = -i \left[\hat{\cal H}, {\hat S}(t-t_0) {\hat \rho}(t)
{\hat S}^\dagger(t-t_0)\right] = -i \left[\hat{\cal H},{\hat
\rho}^{i}(t) \right] \ ,
\end{equation}
where $ \hat{S}(t-t_0) = \hat{U}_\circ(t-t_0)
\hat{U}^\dagger(t-t_0) $ is the unitary transformation connecting
the time evolution of the density-matrix operator in the
Schr\"odinger and interaction pictures. This clearly shows that
the initial quantum-mechanical correlations propagate from $t_0$
to $t$ via the interaction-free dynamics described by the
density-matrix operator written in the interaction picture. As we
shall see, the above quantum-correlation operator is responsible
for a number of purely quantum-mechanical phenomena, like
Hartree-Fock single-particle renormalizations and coherent phonon
effects.\cite{RMP}

The general solution of Eq.~(\ref{LvN-eff_L}) is of the form:
\begin{widetext}
\begin{equation}\label{gen-sol}
\hat\rho(t) = T\left[e^{\int_{t_0}^t \tilde{\cal L}(t')
dt'}\right] \hat\rho(t_0) + \int_{t_0}^t
T\left[e^{\int_{t'}^t \tilde{\cal L}(t'') dt''}\right] \hat C(t')
dt' \ ,
\end{equation}
\end{widetext}
where $T\left[\dots\right]$ is the usual time- or
chronological-ordering operator.\cite{KB,HJ}

At this point a few comments are in order. So far, the only
approximation introduced in our theoretical description is the
adiabatic decoupling between free carrier evolution and various
many-body interactions; this leads to a significant modification
of the system dynamics: while the exact quantum-mechanical
evolution in (\ref{exact}) corresponds to a fully reversible and
isoentropic unitary transformation, the instantaneous
double-commutator structure in (\ref{Ltilde}) describes, in
general, a non-reversible (i.e., non unitary) dynamics [see
Eq.~(\ref{gen-sol})] characterized by energy relaxation and
dephasing; it follows that the system quantum entropy in
(\ref{entropy}) is no more a constant. At this level of
description this behavior is totally nonphysical, clearly showing
the potential failure and intrinsic limitations of the Markov
approximation. However, as discussed below (see
Sect.~\ref{ss-RD}), the Markov limit previously introduced is
usually employed together with a reduced description of the
system, for which such irreversible dynamics is physically
justified.

Let us finally focus on the nature of the effective Liouville superoperator in
(\ref{Ltilde}). As stressed before, this is the sum of a single-commutator term plus a double-commutator
contribution. In the absence of carrier-carrier as well as carrier-quasiparticle interactions, i.e.,
$\hat H' = 0$, the second term vanishes and the system undergoes a reversible unitary transformation
induced by the single-commutator term, which preserves the trace and the positive character of our
density-matrix operator $\hat\rho$. In contrast, the perturbation Hamiltonian $\hat H'$ within the
Markov limit previously introduced will induce, in general, a non-unitary evolution.
Since any effective Liouville superoperator should describe correctly the time evolution of
$\hat\rho$
and since the latter, by definition, needs to be trace-invariant and positive-definite at any time,
it is important to determine if ---and under which conditions--- the superoperator $\tilde{\cal L}$
fulfills this two basic requirements.

As far as the first issue is concerned, recalling that the trace
of a commutator is always equal to zero and taking the trace of
Eq.~(\ref{LvN-eff_L}), it is easy to verify that the
time-derivative of the trace of $\hat\rho$ is equal to zero, i.e.,
that our effective dynamics is trace-preserving.

Let us now discuss the possible positive-definite character of
$\hat\rho$. In general, our effective Liouville superoperator does
not ensure that for any initial condition the density-matrix
operator will be positive-definite at any time. Indeed, it is possible to show that the
double-commutator structure in (\ref{Ltilde}) can be rewritten in
terms of a single-commutator structure (renormalizing the free
Hamiltonian $\hat H_\circ$) and of double commutators of the form:
\begin{equation}\label{LDC}
L\left(\hat{\rho}\right) = -\left[\hat{\cal A}, \left[\hat{\cal A}, \hat\rho\right]\right]\ .
\end{equation}
Each of the latter represents a particular case of the so called
Lindblad superoperators, which are known to
describe completely-positive (CP) maps, thus preserving the
positive character of our density-matrix operator. However, our
Liouville superoperator can be written in terms of the difference
of the Lindblad superoperators in (\ref{LDC}), which in general is
not Lindblad-like.

Since our primary goal is the investigation of quantum-transport
phenomena, we shall focus on the steady-state solution of
Eq.~(\ref{LvN-eff_L}). It is easy to verify that the identity
operator, properly normalized, $\hat\rho(t) \propto \hat {\cal I}$
is the stationary solution we are looking for. A closer inspection
of Eq.~(\ref{gen-sol}) reveals that for any positive-definite and
uncorrelated initial state ($\hat C = 0$) and for a Liouville
superoperator $\tilde{\cal L}$ with a non-positive eigenvalue
spectrum, in the limit $t \to \infty$ the density matrix $\hat
\rho$ will reduce to the identity operator previously mentioned.
As anticipated, this clearly shows that within such approximation
scheme our effective dynamics describes a sort of
decoherence/dephasing, since possible non-diagonal terms of the
density matrix will vanish on the long-time scale. This is again
an artefact of the Markov limit.

Let us finally discuss the physical meaning of the steady-state
solution $\hat\rho(t \to \infty) \propto \hat {\cal I}$. Within
our noninteracting carrier-plus-quasiparticle basis $\lambda$ we
have:
$
\rho_{\lambda_1\lambda_2}(t \to \infty) \propto
\delta_{\lambda_1\lambda_2}
$.
This tells us that, physically speaking, the steady-state solution of our transport equation corresponds
to an equally-probable population of all the microscopic states $\lambda$ without any interstate
quantum coherence ($\rho_{\lambda_1 \neq \lambda_2} = 0$).
This scenario is typical of the present global (carrier + quasiparticle) description; in contrast,
within a reduced description (of the carrier subsystem only) the steady-state solution differs from
the identity operator, since in this case the trace over non-relevant degrees of freedom will translate
into a thermal weight over our electronic states (see Sect.~\ref{ss-RD}).

\subsection{Generalized scattering superoperator}\label{ss-GSS}

Let us now evaluate the explicit form of the scattering
superoperator $\Gamma$ in (\ref{Gamma}). The effective
Liouville-von Neumann equation in (\ref{LvN-eff_L}) can be easily
rewritten in the noninteracting-states basis $\{\vert \lambda
\rangle\}$ previously introduced:
\begin{widetext}
\begin{eqnarray}\label{LvN-eff-lambda}
{d\rho_{\lambda_1\lambda_2} \over dt} &=&
{\epsilon_{\lambda_1}-\epsilon_{\lambda_2} \over i\hbar}
\rho_{\lambda_1\lambda_2} + C_{\lambda_1\lambda_2} +
\sum_{\lambda'_1\lambda'_2}
\Gamma_{\lambda_1\lambda_2,\lambda'_1\lambda'_2}
\rho_{\lambda'_1\lambda'_2} \ .
\end{eqnarray}
\end{widetext}
In order to derive the explicit form of the superoperator matrix elements
$\Gamma_{\lambda_1\lambda_2,\lambda'_1\lambda'_2}$,
let us expand the double commutator in (\ref{Gamma}):
\begin{widetext}
\begin{eqnarray}\label{Gamma-lambda1}
\sum_{\lambda'_1\lambda'_2}
\Gamma_{\lambda_1\lambda_2,\lambda'_1\lambda'_2}
\rho_{\lambda'_1\lambda'_2} &=& \left[\hat{\cal H},
\left[\hat{\cal K}, {\hat \rho}\right]\right]_{\lambda_1\lambda_2}
= \sum_{\lambda'_1\lambda'_2} \left( {\cal
H}_{\lambda_1\lambda'_1} \rho_{\lambda'_1\lambda'_2} {\cal
K}_{\lambda'_2\lambda_2} \right. + \left. {\cal
K}_{\lambda_1\lambda'_1} \rho_{\lambda'_1\lambda'_2} {\cal
H}_{\lambda'_2\lambda_2} \right)
\nonumber \\
&-& \sum_{\lambda'_1\lambda'_2} \left( {\cal
H}_{\lambda_1\lambda'_1} {\cal K}_{\lambda'_1\lambda'_2}
\rho_{\lambda'_2\lambda_2} \right. + \left.
\rho_{\lambda_1\lambda'_1} {\cal K}_{\lambda'_1\lambda'_2} {\cal
H}_{\lambda'_2\lambda_2}\right) \ .
\end{eqnarray}
\end{widetext}
As we can see, the scattering operator $\Gamma$ can be written as
the difference of the following ``in-'' and ``out-scattering''
terms:
%\begin{equation}\label{Gamma-lambda2}
%\Gamma_{\lambda_1\lambda_2,\lambda'_1\lambda'_2} =
%\Gamma^{\rm in}_{\lambda_1\lambda_2,\lambda'_1\lambda'_2} -
%\Gamma^{\rm out}_{\lambda_1\lambda_2,\lambda'_1\lambda'_2}
%\end{equation}
%with
\begin{equation}\label{in1}
\Gamma^{\rm in}_{\lambda_1\lambda_2,\lambda'_1\lambda'_2} =
{\cal H}_{\lambda_1\lambda'_1} {\cal K}_{\lambda'_2\lambda_2} +
{\cal K}_{\lambda_1\lambda'_1} {\cal H}_{\lambda'_2\lambda_2}
\end{equation}
%and
\begin{widetext}
\begin{eqnarray}\label{out1}
\Gamma^{\rm out}_{\lambda_1\lambda_2,\lambda'_1\lambda'_2} =
\sum_{\lambda''} {\cal H}_{\lambda_1\lambda''} {\cal
K}_{\lambda''\lambda'_1} \delta_{\lambda_2\lambda'_2}  +
\sum_{\lambda''} \delta_{\lambda_1\lambda'_1} {\cal
K}_{\lambda'_2\lambda''} {\cal H}_{\lambda''\lambda_2}
 \ .
\end{eqnarray}
\end{widetext}

For the case of a time-independent perturbation ${\hat H}' = \hbar
\hat{\cal H}$, the operator $\hat{\cal K}$ can be rewritten as:
\begin{equation}\label{calK3}
\hat{\cal K} = \int_0^{t-t_0} d\tau
{\hat U}^{ }_\circ(\tau) \hat{\cal H} {\hat U}^\dagger_\circ(\tau) \ .
\end{equation}
Taking into account that within the $\lambda$-representation the noninteracting evolution
operator ${\hat U}_\circ$ is simply given by
\begin{equation}\label{U0-lambda}
U^{\lambda\lambda'}_\circ(\tau) = e^{\epsilon_\lambda \tau \over i\hbar} \delta_{\lambda\lambda'} \ ,
\end{equation}
the matrix elements of the operator $\hat{\cal K}$ in
(\ref{calK3}) result to be:
\begin{equation}\label{calK3-lambda}
{\cal K}_{\lambda\lambda'} = 2\pi {\cal H}_{\lambda\lambda'} {\cal D}_{\lambda\lambda'}
\end{equation}
with
\begin{equation}\label{calD}
{\cal D}_{\lambda\lambda'} = {1 \over 2\pi} \int_0^{t-t_0}
e^{(\epsilon_\lambda-\epsilon_{\lambda'})\tau \over i \hbar} d\tau =
{\cal D}^*_{\lambda'\lambda}
\ .
\end{equation}
By inserting the above result into Eqs.~(\ref{in1})-(\ref{out1})
and recalling that ${\cal H}_{\lambda\lambda'} \equiv
{H'_{\lambda\lambda'} \over \hbar} = {\cal
H}^*_{\lambda'\lambda}$, we finally obtain:
\begin{widetext}
\begin{equation}\label{in2}
\Gamma^{\rm in}_{\lambda_1\lambda_2,\lambda'_1\lambda'_2} =
{2\pi \over \hbar^2} \left(
H'_{\lambda_1\lambda'_1} H^{\prime *}_{\lambda_2\lambda'_2} {\cal D}^*_{\lambda_2\lambda'_2} +
{\cal D}_{\lambda_1\lambda'_1} H'_{\lambda_1\lambda'_1} H^{\prime *}_{\lambda_2\lambda'_2}\right) \ ;
\end{equation}
\begin{equation}\label{out2}
\Gamma^{\rm out}_{\lambda_1\lambda_2,\lambda'_1\lambda'_2} =
{2\pi \over \hbar^2} \sum_{\lambda''}
\left(
H^{\prime *}_{\lambda''\lambda_1} H'_{\lambda''\lambda'_1} {\cal D}_{\lambda''\lambda'_1}
\delta_{\lambda_2\lambda'_2}
+
\delta_{\lambda_1\lambda'_1}
{\cal D}^*_{\lambda''\lambda'_2} H^{\prime *}_{\lambda''\lambda'_2} H'_{\lambda''\lambda_2}
\right) \ .
\end{equation}
\end{widetext}
%We finally stress that,
In general, the scattering superoperator $\Gamma$ is a function of
time; however, in the limit $t_0 \to -\infty$ ---also called
``completed-collision limit''~\cite{RBJ,QMC}--- the function
${\cal D}$ in (\ref{calD}) becomes time-independent:
\begin{equation}\label{calD2}
{\cal D}^{-\infty}_{\lambda\lambda'} = {1 \over 2\pi}
\int_0^\infty e^{(\epsilon_\lambda-\epsilon_{\lambda'})\tau \over i \hbar} d\tau \ .
\end{equation}
It follows that in this limit the operator ${\cal K}$ as well as
the superoperators $\Gamma$ and $\tilde{\cal L}$ are also
time-independent. In this case there is no need for the
time-ordering operator $T$ in (\ref{gen-sol}). Moreover, the real
part of the function ${\cal D}^{-\infty}$ in (\ref{calD2}) gives
the well-known energy-conserving Dirac delta function, i.e.,
\begin{equation}\label{calD3}
{\cal D}^{-\infty}_{\lambda\lambda'} = {\hbar \over 2}
\delta(\epsilon_\lambda-\epsilon_{\lambda'}) + {\cal
R}_{\lambda\lambda'} \ ,
\end{equation}
while the imaginary part ---denoted by ${\cal
R}_{\lambda\lambda'}$--- describes, in general,
energy-renormalization effects. Within the validity limits of the
present Markov treatment, such renormalization effects can be
safely neglected: if, as requested, the perturbation Hamiltonian
is small compared to the noninteracting one, then the resulting
energy-level renormalization is small compared to the
noninteracting energy levels $\epsilon_\lambda$.

At this point a few comments on the evaluation of the time
integral in (\ref{calD2}) are in order. Indeed, it is well known
that the limit $t_\circ \to -\infty$ needs to be performed
properly; more specifically, this is realized by adding to the
energy difference an infinitesimally small imaginary part, which
ensures the convergence of the time integration. A qualitative
---but not rigorous--- interpretation of such mathematical
prescription is based on the so-called ``adiabatic switching-on''
procedure:\cite{Mahan} The idea is that, starting from $t =
-\infty$, the interaction mechanism/Hamiltonian is slowly or
adiabatically switched on. By employing the nonequilibrium Green's
function formalism, it is possible to show~\cite{HJ} that such
imaginary part is not an artificial ingredient: it corresponds to
the imaginary part of the electron self-energy, thus describing
the finite life-time of our electronic states due to all relevant
interaction mechanisms. A proper account of such effect ---not
relevant in the present discussion--- leads to apparent violations
of the energy-conserving transitions predicted by the Dirac delta
function in ({\ref{calD3}), the so-called ``collisional
broadening''.

As previously recalled, it is imperative to establish if ---and
under which conditions--- the scattering superoperator in
(\ref{Gamma-lambda1}) preserves the positive-definite nature of
our density-matrix operator $\rho$. As discussed in
App.~\ref{a-PDC}:
\begin{itemize}
\item[(i) ]
contrary to the semiclassical or Boltzmann dynamics (see
Sect.~\ref{ss-SL}), the effective Liouville superoperator
previously identified does not correspond to a so-called ``CP
map'', i.e., it does not preserve, in
general, the positive-definite character of our density-matrix
operator;
\item[(ii) ]
its eigenvalue spectrum, i.e.,
\begin{equation}\label{EP}
\tilde{\cal L}\left(\hat \rho\right) = \Lambda \hat \rho \ ,
\end{equation}
always contains the $\Lambda = 0$ eigenvalue, which corresponds to the steady-state
transport solution;
\item[(iii) ]
in the ``small-coupling limit'' it is possible to show that the steady-state density-matrix
operator ---corresponding to the $\Lambda = 0$ eigenvalue--- is always positive definite, i.e.,
\begin{equation}\label{PD}
\hat\rho^{\Lambda = 0} = \sum_{\bar{\lambda}} P_{\bar{\beta}}
\vert {\bar{\beta}} \rangle \langle {\bar{\beta}} \vert \ ,
\end{equation}
where the basis states $\vert \bar{\beta} \rangle$ are the
eigenvectors of $\hat{\rho}^{\Lambda = 0}$, and $P_{\bar{\beta}}$
is a (non-negative) probability distribution.
\end{itemize}

\subsection{Semiclassical limit}\label{ss-SL}

The well-known semiclassical or Boltzmann transport
theory~\cite{MC} can be easily derived from the quantum-transport
formulation presented so far, by introducing the so-called
diagonal or semiclassical limit. The latter corresponds to
neglecting all non-diagonal density-matrix elements (and therefore
any quantum-mechanical phase coherence between the generic states
$\lambda_1$ and $\lambda_2$), i.e.,
\begin{equation}\label{SL1}
\rho_{\lambda_1\lambda_2} = f_{\lambda_1} \delta_{\lambda_1\lambda_2} \ ,
\end{equation}
where the diagonal elements $f_\lambda$ describe the semiclassical
distribution function over the noninteracting basis states
$\lambda$.

By introducing the above semiclassical density matrix into
Eq.~(\ref{LvN-eff-lambda}) for the diagonal elements ($\lambda =
\lambda_1 = \lambda_2$),
%we get:
%\begin{equation}\label{SL2}
%{d f_\lambda \over dt} =
%\sum_{\lambda'} \Gamma_{\lambda\lambda,\lambda'\lambda'} f_{\lambda'} \ .
%\end{equation}
%
%; by rewriting the diagonal
%matrix elements of the correlation operator in (\ref{C}) within
%the semiclassical limit we get:
%\begin{equation}\label{C-SL}
%C_{\lambda\lambda} =
%-i\left(
%{\cal H}_{\lambda\lambda} f_\lambda(t_0) -
%f_\lambda(t_0) {\cal H}_{\lambda\lambda}\right) = 0 \ .
%\end{equation}
%
and inserting %into Eq.~(\ref{SL2})
the explicit form of the elements $\Gamma_{\lambda \lambda}$ of
the scattering operator in the limit $t_0 \to -\infty$ [see
Eq.~(\ref{calD2})],
%---as sum of its in- and out-contributions in Eqs.~(\ref{in2}) and (\ref{out2})---
we finally obtain the usual form of the Boltzmann transport
equation written in our basis states:
\begin{equation}\label{SL3}
{d f_\lambda \over dt} =
\sum_{\lambda'} \left(
P_{\lambda\lambda'} f_{\lambda'} - P_{\lambda'\lambda} f_\lambda
\right)\ ,
\end{equation}
where
\begin{equation}\label{P}
P_{\lambda\lambda'} = P_{\lambda'\lambda} =
{2\pi \over \hbar} |H'_{\lambda'\lambda}|^2 \delta\left(\epsilon_{\lambda'}-\epsilon_\lambda\right)
\end{equation}
are the semiclassical scattering rates given by the well-known
Fermi's golden rule. In addition to the square of the interaction
matrix element $H'_{\lambda'\lambda}$, they contain the
energy-conserving Dirac delta function:
\begin{equation}\label{delta}
\delta\left(\epsilon_{\lambda'}-\epsilon_\lambda\right) =
{
{\cal D}^{-\infty}_{\lambda\lambda'}+{\cal D}^{-\infty}_{\lambda'\lambda}
\over \hbar
}
=
{1 \over 2\pi} \int_{-\infty}^\infty e^{i(\epsilon_{\lambda'}-\epsilon_{\lambda})\tau } d\tau \ .
\end{equation}
Within the semiclassical limit the free-rotation term in
(\ref{LvN-eff-lambda}) vanishes, and the same applies to the
quantum-correlation contributions $C_{\lambda\lambda}$.

Our analysis shows that the quantum-transport equation in
(\ref{LvN-eff-lambda}) can be regarded as the quantum-mechanical
generalization of the Boltzmann equation in (\ref{SL3}). Indeed,
the in- and out-scattering superoperators in (\ref{Gamma-lambda1})
are the quantum-mechanical generalizations of the standard in- and
out-scattering terms entering the Boltzmann collision operator in
(\ref{SL3}).

As a confirmation of the fact that the Markov approximation leads
to a totally nonphysical non-reversible (i.e., non-unitary) system
evolution, it is possible to show that the system entropy $S$ in
Eq.~(\ref{entropy}) is a non-decreasing function of time.

We stress that, contrary to the usual semiclassical transport theory, the Boltzmann-like
equation in (\ref{SL3}) describes a scattering dynamics within the whole $\lambda = \nalpha,\nqu$
space, i.e., the generic scattering probability $P_{\lambda\lambda'}$ in (\ref{P}) describes
a transition from the state $\nalpha,\nqu$ to the state
$\nalphap,\nqup$. In other words, so far no reduction procedure to the $\alpha$-subsystem has
been performed; it follows that for a given transition of the $\alpha$ subsystem
($\nalpha \to \nalphap$), a corresponding transition ($\nqu \to \nqup$) of the quasiparticle
subsystem will also take place. This explains why, contrary to the usual Boltzmann theory,
the scattering probabilities in (\ref{P}) are symmetric, i.e., invariant under time reversal:
$P_{\lambda\lambda'} = P_{\lambda'\lambda}$; moreover, the Dirac delta function in (\ref{P})
leads to the conservation of the total energy of the system.

A second important remark is that, contrary to the non-diagonal density-matrix description
previously introduced, the Markov limit combined with the semiclassical or diagonal approximation
in (\ref{SL1}) ensures that at any time $t$ our semiclassical distribution function $f_\lambda$
is always positive-definite (see App.~\ref{a-PDC}).

Let us finally discuss the steady-state solution of the Boltzmann transport equation in (\ref{SL3}).
From the detailed-balance principle, i.e.,
\begin{equation}\label{DBP}
P_{\lambda\lambda'} f_{\lambda'} = P_{\lambda'\lambda} f_\lambda \ ,
\end{equation}
and considering that the semiclassical scattering rates in (\ref{P}) are symmetric
($P_{\lambda\lambda'} = P_{\lambda'\lambda}$), we get:
\begin{equation}\label{uni-dis}
{f_{\lambda'} \over f_\lambda} = {P_{\lambda\lambda'} \over
P_{\lambda'\lambda}} = 1 \ .
\end{equation}
Exactly as in the quantum-mechanical case, the steady-state
solution corresponds to a uniform distribution over the
noninteracting carrier-plus-quasiparticle states:
%$\lambda$
$f_\lambda \propto \delta_{\lambda \lambda}$.
%, which coincides with the identity-operator solution in (\ref{calI}).
As discussed in Sect.~\ref{ss-RD}, this is not the case when our kinetic description is
reduced to the $\alpha$ subsystem only.

\subsection{Reduced description}\label{ss-RD}

As discussed in Sect.~\ref{ss-DMF}, the average value of any given
physical quantity $A$ can be easily expressed in terms of the
density-matrix operator $\hat\rho$ according to
Eq.~(\ref{average1}). In the study of charge-transport phenomena
in semiconductor nanostructures, most of the physical quantities
of interest depend on the electronic-subsystem coordinates
$\alpha$ only (carrier drift velocity, total electronic energy,
carrier-carrier correlation function, etc.), i.e.,
\begin{equation}\label{average4}
A_{\nalpha,\nqu;\nalphap,\nqup} = A_{\nalpha,\nalphap} \delta_{\nqu,\nqup}
\ .
\end{equation}
In this case it is convenient to write %Eq.~(\ref{average3}) as
\begin{equation}\label{average5}
A = \sum_{\lambda \lambda'} A_{\lambda \lambda'} \rho_{\lambda'
\lambda} = \sum_{\nalpha,\nalphap} A_{\nalpha,\nalphap}
\rho^c_{\nalphap,\nalpha} \ ,
\end{equation}
where
\begin{equation}\label{average6}
\rho^c_{\nalpha,\nalphap} = \sum_{\nqu} \rho_{\nalpha,\nqu;\nalphap,\nqu}
\end{equation}
is the so-called reduced or electronic density matrix. Equation
(\ref{average6}) can be also written in an operatorial form as:
\begin{equation}\label{average7}
\hat\rho^c = \rmtr\left\{\hat\rho\right\}_\nqu \ ,
\end{equation}
which shows that the electronic density-matrix operator
$\hat\rho^c$ is obtained by performing a trace operation over the
quasi-particle coordinates $\qu$. Since $\hat\rho^c$ is the only
quantity entering the evaluation of the average value in
(\ref{average5}), it is desirable to derive a corresponding
equation of motion for the reduced density-matrix operator.
Combining Eqs.~(\ref{LvN-eff_L}) and (\ref{average7}) we get:
\begin{equation}\label{LvN-eff_Lc}
{d{\hat \rho}^c \over dt} = \rmtr\left\{\tilde{\cal L}\left({\hat
\rho}\right)\right\}_\nqu + \rmtr\left\{\hat C\right\}_\nqu \ .
\end{equation}
In general, the trace over the quasiparticle coordinates does not
commute with the Liouville superoperator $\tilde{\cal L}$ in
(\ref{Ltilde}), which does not allow to obtain a closed equation
of motion for the reduced density-matrix operator $\hat\rho^c$.
This clearly does not apply when the interaction Hamiltonian is a
function of the carrier coordinates only, e.g., carrier-carrier,
carrier-impurity, etc. In contrast, in the presence of
carrier-quasiparticle coupling additional approximations are
needed. In order to get a closed equation of motion for the
reduced density-matrix operator, the typical assumption is to
consider the quasiparticle subsystem as characterized by a huge
number of degrees of freedom (compared to the subsystem $\alpha$).
In other words this amounts to say that the $\qu$ subsystem has an
infinitely high heat capacity, i.e., it behaves as a thermal bath;
this allows to consider the quasiparticle subsystem always in
thermal equilibrium, i.e., not significantly perturbed by the
carrier subsystem $\alpha$. Within such approximation scheme, the
global ($\alpha + \qu$) density-matrix operator $\hat\rho$ can be
written as the product of the equilibrium density-matrix operator
for the quasiparticle subsystem $\hat\rho^{qp}$ and the reduced
density-matrix operator $\hat\rho^c$:
\begin{equation}\label{fact1}
\hat\rho = \hat\rho^c \otimes \hat\rho^{qp}\ , \qquad
\hat\rho^{qp} = { e^{-{\hat H_\circ^{qp} \over k_{\rm B} T}} \over
\rmtr\left\{e^{-{\hat H_\circ^{qp} \over k_{\rm B} T}}\right\} } \
.
\end{equation}
The corresponding matrix elements within our basis states $\lambda = \nalpha,\nqu$ are then given by:
\begin{equation}\label{fact2}
\rho_{\lambda\lambda'} = \rho^c_{\nalpha \nalphap} f^{qp}_{\nqu} \delta_{\nqu \nqup}
\end{equation}
with
\begin{equation}\label{fact2-bis}
f^{qp}_{\nqu} = { e^{-{\epsilon_{\nqu} \over k_{\rm B} T}} \over
\sum_{\nqu} e^{-{\epsilon_{\nqu} \over k_{\rm B} T}} } \ .
\end{equation}
By inserting these density-matrix elements into Eq.~(\ref{LvN-eff-lambda}) and performing the trace
over the quasiparticle coordinates, it is easy to get the following effective Liouville-von Neumann
equation for the reduced density matrix $\rho^c$:
\begin{widetext}
\begin{equation}\label{fact3}
{d\rho^c_{\nalphaone \nalphatwo} \over dt} =
{\epsilon_{\nalphaone}-\epsilon_{\nalphatwo} \over i\hbar}
\rho^c_{\nalphaone \nalphatwo} + C^c_{\nalphaone \nalphatwo} +
\sum_{\nalphaonep \nalphatwop} \Gamma^c_{\nalphaone \nalphatwo,
\nalphaonep \nalphatwop} \rho^c_{\nalphaonep \nalphatwop}
\end{equation}
\end{widetext}
with
\begin{widetext}
\begin{equation}\label{fact4}
\Gamma^c_{\nalphaone \nalphatwo, \nalphaonep \nalphatwop} = \sum_{\nqu \nqup}
\Gamma_{\nalphaone\nqu,\nalphatwo\nqu;\nalphaonep\nqup,\nalphatwop\nqup} f^{qp}_{\nqup}
\end{equation}
\end{widetext}
and
\begin{equation}\label{fact5}
C^c_{\nalphaone\nalphatwo} = \sum_{\nqu}
C_{\nalphaone\nqu,\nalphatwo\nqu} \ .
\end{equation}
By denoting with $\Gamma^c$ the effective scattering superoperator defined by the matrix elements
in (\ref{fact4}) ---acting on the $\alpha$ Hilbert subspace only---
the new effective Liouville superoperator (i.e., traced over the $\qu$ coordinates) is given by:
\begin{equation}\label{fact6}
\tilde{\cal L}^c\left({\hat \rho^c}\right) =
{1 \over i\hbar} \left[{\hat H}^c_\circ, {\hat \rho^c}\right] +
\Gamma^c\left(\hat\rho^c\right)\ .
\end{equation}
%The analysis presented so far shows that, by assuming the
%factorized solution in (\ref{fact1}), also in the presence of
%carrier-quasiparticle interaction ---i.e., carrier-phonon,
%carrier-plasmon, etc.--- one is able to derive a closed equation
%of motion for the reduced density-matrix operator $\hat\rho^c$
%formally identical to Eq.~(\ref{LvN-eff_L-CC}).
%However, the new effective Liouville superoperator in (\ref{fact6})
The latter, however, does not contain the double-commutator
structure previously discussed; This aspect will be more
extensively addressed in the following.

For all relevant carrier-quasiparticle interaction mechanisms in semiconductor nanostructures
---e.g., carrier-phonon, carrier-plasmon, etc.--- the perturbation Hamiltonian $\hat H'$ can be
written as:
\begin{equation}\label{H_c-qp}
\hat H' = \hbar \hat{\cal H} = \hbar \sum_\qu \left(\hat{\cal H}_\qu
\hat b^{ }_\qu + \hat{\cal H}^\dagger_\qu \hat b^\dagger_\qu \right)
%= \hat{H}^{ab} + \hat{H}^{em} \ .
\end{equation}
Here $\hat{\cal H}_\qu = \hat{\cal H}^\dagger_{-\qu}$ are electronic operators (parameterized by
the quasiparticle wavevector $\qu$) acting on the $\alpha$ subsystem only.
The two terms in (\ref{H_c-qp}) ---corresponding to quasiparticle destruction and creation---
describe quasiparticle absorption and emission processes.

Let us consider again the definition of the effective
carrier-quasiparticle scattering superoperator $\Gamma^c$ in
(\ref{fact4}) written in operatorial form, i.e.,
\begin{equation}\label{Gamma_c-2}
\Gamma^c\left({\hat \rho^c}\right) = - \rmtr\left\{\left[\hat{\cal
H}, \left[\hat{\cal K}, \hat\rho^c
\hat\rho^{qp}\right]\right]\right\}_\nqu \ ,
\end{equation}
where %Introducing into Eq.~(\ref{Gamma_c-1})
the dou\-ble-com\-mu\-ta\-tor form in (\ref{Gamma}) has been
introduced.

By inserting into Eqs.~(\ref{calK3}) and (\ref{Gamma_c-2}) the
explicit form of the carrier-quasiparticle Hamiltonian in
(\ref{H_c-qp}) and using the bosonic commutation relations $[\hat
b^{ }_\qu, \hat b^\dagger_{\qu'}] = \delta_{\qu\qu'}$, we obtain
an explicit form of the effective carrier-quasiparticle scattering
superoperator in (\ref{Gamma_c-2}). More specifically we get:
\begin{equation}\label{Gamma_c-3}
\Gamma^c\left({\hat \rho^c}\right) = -\sum_{\qu \pm}
\left(N_\qu+{1 \over 2} \pm {1 \over 2}\right)
\left[
\hat{\cal H}_\qu ,
\hat{\cal K}^{\pm}_\qu {\hat \rho}^c
-
{\hat \rho}^c \hat{\cal K}^\mp_\qu \right]
\end{equation}
with
\begin{equation}\label{calK_c}
\hat{\cal K}^\pm_\qu =
\int_0^{t-t_0} d\tau e^{{\hat H}^c_\circ \tau
\over i \hbar} \hat{\cal H}^\dagger_\qu e^{-{{\hat H}^c_\circ \tau
\over i \hbar}} e^{\pm{\epsilon_\qu \tau \over i \hbar}}\ .
\end{equation}
Here
\begin{equation}\label{B-H}
N_\qu = \rmtr\left\{\hat b^\dagger_\qu \hat b^{ }_\qu
\hat\rho^{qp} \right\} = { 1 \over e^{{\epsilon_\qu \over k_B T}}
- 1 }
\end{equation}
denotes the equilibrium average occupation number for the
quasiparticle state $\qu$. As we can see, for each quasiparticle
state $\qu$ we have two contributions ($\pm$) describing
quasiparticle emission and absorption.

The effective scattering superoperator $\Gamma^c$ in
(\ref{Gamma_c-3}) does not exhibit the double-commutator structure
typical of the global description [see Eq.~(\ref{Gamma})]. Indeed,
denoting with
\begin{equation}\label{K1K2}
\hat{\cal K}^\pm_1 =
{1 \over 2} \left(\hat{\cal K}^\pm_\qu +\hat{\cal K}^\mp_\qu \right) \ ,\qquad
\hat{\cal K}^\pm_2 =
{1 \over 2} \left(\hat{\cal K}^\pm_\qu - \hat{\cal K}^\mp_\qu \right) \ ,
\end{equation}
the superoperator $\Gamma^c$ in (\ref{Gamma_c-3}) can be also expressed as:
\begin{widetext}
\begin{equation}\label{Gamma_c-4}
\Gamma^c\left({\hat \rho^c}\right) = -\sum_{\qu \pm}
\left(N_\qu +{1 \over 2} \pm {1 \over 2}\right)
\left(
\left[\hat{\cal H}_\qu,
\left[\hat{\cal K}^\pm_1,{\hat \rho}^c\right]\right]
+
\left[\hat{\cal H}_\qu,\left\{\hat{\cal K}^\pm_2,{\hat \rho}^c\right\}\right] \right) \ .
\end{equation}
\end{widetext}
As anticipated, the scattering superoperator involves again a double-commutator term, but we
have also a com\-mu\-ta\-tor-anti\-commutator
contribution.

To better underline the physical role played by the above double-com\-mu\-ta\-tor versus
com\-mu\-ta\-tor-an\-ti\-com\-mu\-ta\-tor contributions, let us recall a simplified model
usually invoked to qualitatively describe the quan\-tum-me\-cha\-ni\-cal e\-vo\-lu\-ti\-on of open
sys\-tems, i.e., sub\-sys\-tems in\-te\-rac\-ting with their en\-vi\-ron\-ment.
Within the Schr\-\"o\-din\-ger pic\-tu\-re, the latter are typically treated by adding to the system
Hamiltonian ${\hat H}$ an antihermitian part (imaginary potential) ${\hat V}^{\rm env}$ describing the
system-environment coupling:
\begin{equation}\label{Schr-open}
i\hbar {d \over dt} \vert \psi \rangle = \left({\hat H} + {\hat  V}^{\rm env}\right) \vert \psi \rangle \ .
\end{equation}
Starting from the above modified Schr\"odinger equation, it is easy to obtain a corresponding version of
the Liouville-von Neumann equation in (\ref{LvN1}):
\begin{equation}\label{LvN1-open}
{d{\hat \rho} \over dt} =
{1 \over i\hbar} \left[{\hat H}, {\hat \rho}\right]
+
{1 \over i\hbar} \left\{ {\hat V}^{\rm env}, {\hat \rho}\right\}
\ .
\end{equation}
In addition to the commutator-like dynamics typical of a closed
system, we deal with an anticommutator term, describing
dissipation induced by the system-environment coupling. Indeed,
contrary to the closed dynamics in (\ref{LvN1}), the latter leads
to a non-reversible dynamics. Such a simplified model is known to
be highly nonphysical, since it does not preserve the trace of the
density-matrix operator; however, it clearly shows how the
commutator structure is intimately related to a closed evolution,
while anticommutator terms always describe dissipation processes.

Contrary to the above simplified model, the scattering
superoperator in (\ref{Gamma_c-3}) ---in view of its
outer-commutator structure--- is trace-preserving. However, the
presence of the com\-mu\-ta\-tor-anti\-com\-mu\-ta\-tor
con\-tri\-bu\-tion is a clear fingerprint of carrier-quasiparticle
dissipation phenomena leading to genuine
energy-relaxation/dephasing processes.

As far as the correlation term in (\ref{fact5}) is concerned, it
is easy to verify that the latter vanishes for the linear-coupling
carrier-quasiparticle Hamiltonian in (\ref{H_c-qp}).

In summary, within the approximation scheme considered so far we get the following effective equation
of motion for the reduced density-matrix operator $\hat\rho^c$:
\begin{equation}\label{LvN-eff_L-cqp}
{d{\hat \rho}^c \over dt} =
\tilde{\cal L}^c\left({\hat \rho^c}\right)
\end{equation}
with $\tilde{\cal L}^c$ defined in~(\ref{fact6}). In the limit
$t_0 \to -\infty$, the effective Liouville superoperator in
(\ref{fact6}) becomes time-independent, and the general solution
of the homogeneous equation in (\ref{LvN-eff_L-cqp}) is of the
form:
\begin{equation}\label{gen-sol-2}
\hat\rho^c(t) = e^{\tilde{\cal L}^c (t-t_0)} \hat\rho^c(t_0) \ .
\end{equation}
Again, contrary to the isoentropic and fully-re\-ver\-si\-ble unitary evolution in (\ref{exact}),
the instantaneous dou\-ble-com\-mu\-ta\-tor plus com\-mu\-ta\-tor-an\-ti\-com\-mu\-ta\-tor structures
in (\ref{Gamma_c-4}) describe a
non-reversible (i.e., non unitary) dynamics characterized by energy relaxation and dephasing induced
by the carrier-quasiparticle coupling in (\ref{H_c-qp}).

We shall now derive the explicit form of the scattering superoperator $\Gamma^c$ within our
noninteracting-carrier basis $\nalpha$. To this end let us expand the various terms entering
Eq.~(\ref{Gamma_c-3}):
\begin{widetext}
\begin{eqnarray}\label{Gamma-nalpha1}
\sum_{\nalphaonep\nalphatwop}
\Gamma^c_{\nalphaone\nalphatwo,\nalphaonep\nalphatwop}
\rho^c_{\nalphaonep\nalphatwop} &=& \sum_{\qu \pm} \left(N_\qu +{1
\over 2} \pm {1 \over 2}\right) \nonumber \\
\sum_{\nalphaonep\nalphatwop} \left[\left( {\cal
H}^{\qu}_{\nalphaone\nalphaonep} \rho^c_{\nalphaonep\nalphatwop}
{\cal K}^{\qu \mp}_{\nalphatwop\nalphatwo} \right.\right. &+&
\left.\left. {\cal K}^{\qu \pm}_{\nalphaone\nalphaonep}
\rho^c_{\nalphaonep\nalphatwop}
{\cal H}^{\qu}_{\nalphatwop\nalphatwo} \right)\right.\nonumber \\
- \left.\left({\cal H}^{\qu}_{\nalphaone\nalphaonep} {\cal K}^{\qu
\pm}_{\nalphaonep\nalphatwop} \rho^c_{\nalphatwop\nalphatwo}
\right.\right. &+& \left.\left. \rho^c_{\nalphaone\nalphaonep}
{\cal K}^{\qu \mp}_{\nalphaonep\nalphatwop} {\cal
H}^{\qu}_{\nalphatwop\nalphatwo}\right)\right] \ .
\end{eqnarray}
\end{widetext}
As we can see, also in this case the scattering superoperator
$\Gamma^c$ can be regarded as the difference of the following in-
and out-scattering terms:
\begin{widetext}
\begin{equation}\label{in1_c}
\Gamma^{\rm in}_{\nalphaone\nalphatwo,\nalphaonep\nalphatwop} =
\sum_{\qu \pm} \left(N_\qu +{1 \over 2} \pm {1 \over 2}\right)
\left( {\cal H}^{\qu}_{\nalphaone\nalphaonep} {\cal K}^{\qu
\mp}_{\nalphatwop\nalphatwo} + {\cal K}^{\qu
\pm}_{\nalphaone\nalphaonep} {\cal
H}^{\qu}_{\nalphatwop\nalphatwo} \right) \ ,
\end{equation}
\begin{eqnarray}\label{out1_c}
\Gamma^{\rm out}_{\nalphaone\nalphatwo,\nalphaonep\nalphatwop} &=&
\sum_{\nalphas, \qu \pm} \left(N_\qu +{1 \over 2} \pm {1 \over
2}\right)  \left( {\cal H}^{\qu}_{\nalphaone\nalphas} {\cal
K}^{\qu \pm}_{\nalphas\nalphaonep}
\delta_{\nalphatwo\nalphatwop}\right. \nonumber \\ &+& \left.
\delta_{\nalphaone\nalphaonep} {\cal K}^{\qu
\mp}_{\nalphatwop\nalphas} {\cal H}^{\qu}_{\nalphas\nalphatwo}
\right) \ .
\end{eqnarray}
\end{widetext}
As for the case of the global (carrier + quasiparticle)
description presented in Sect.~\ref{ss-GSS}, the matrix elements
${\cal K}^{\qu \pm}_{\nalpha\nalphap}$ of the operator $\hat{\cal
K}^\pm_\qu$ in (\ref{calK_c}) may be expressed in terms of the
matrix elements of the operator $\hat{\cal H}_\qu$:
\begin{equation}\label{calK3-nalpha}
{\cal K}^{\qu \pm}_{\nalpha\nalphap} = 2\pi
{\cal H}^{\qu *}_{\nalphap\nalpha} {\cal D}^{\qu \mp *}_{\nalphap\nalpha}
\end{equation}
with
\begin{equation}\label{calD_c}
{\cal D}^{\qu \pm}_{\nalpha\nalphap} =
{1 \over 2\pi} \int_0^{t-t_0}
e^{(\epsilon_\nalpha-\epsilon_{\nalphap} \pm \epsilon_\qu)\tau \over i \hbar} d\tau
\ .
\end{equation}
Again, in the completed-collision limit ($t_0 \to -\infty$), the real part of
${\cal D}$ provides the energy-conserving Dirac delta function, while its imaginary
part describes carrier-quasiparticle energy-renormalization effects.
By inserting this relation into the above in- and out-scattering rates and recalling that
${\cal H}^{\qu *}_{\nalpha\nalphap} = {\cal H}^{-\qu}_{\nalphap\nalpha}$
and
${\cal D}^{\qu \pm *}_{\nalpha\nalphap} = {\cal D}^{\qu \mp}_{\nalphap\nalpha}$,
we finally get:
\begin{widetext}
\begin{eqnarray}\label{in2_c}
\Gamma^{\rm in}_{\nalphaone\nalphatwo,\nalphaonep\nalphatwop} &=&
{2\pi \over \hbar^2} \sum_{\qu \pm} \left(N_\qu +{1 \over 2} \pm
{1 \over 2}\right) \left( {H}^{\qu}_{\nalphaone\nalphaonep}
{H}^{\qu *}_{\nalphatwo\nalphatwop} {\cal D}^{\qu \pm
*}_{\nalphatwo\nalphatwop} \right. \nonumber \\ &+& \left. {\cal
D}^{\qu \pm}_{\nalphaone\nalphaonep}
{H}^{\qu}_{\nalphaone\nalphaonep} {H}^{\qu
*}_{\nalphatwo\nalphatwop} \right)
\end{eqnarray}
and
\begin{eqnarray}\label{out2_c}
\Gamma^{\rm out}_{\nalphaone\nalphatwo,\nalphaonep\nalphatwop} &=&
{2\pi \over \hbar^2} \sum_{\nalphas, \qu \pm} \left(N_\qu +{1
\over 2} \pm {1 \over 2}\right)  \left( {H}^{\qu
*}_{\nalphas\nalphaone} {H}^{\qu}_{\nalphas\nalphaonep} {\cal
D}^{\qu \pm}_{\nalphas\nalphaonep}
\delta_{\nalphatwo\nalphatwop} \right. \nonumber \\
&& + \left. \delta_{\nalphaone\nalphaonep}
{\cal D}^{\qu \pm *}_{\nalphas\nalphatwop}
{H}^{\qu *}_{\nalphas\nalphatwop}
{H}^{\qu}_{\nalphas\nalphatwo}
\right) \ ,
\end{eqnarray}
\end{widetext}
where $H^\qu_{\nalpha\nalphap} = \hbar {\cal H}^\qu_{\nalpha\nalphap}$.
We stress that the above in- and out-scattering superoperators are linear, i.e.,
$\rho$-independent. As we shall see, this feature ---typical of the present many-electron description
$\nalpha$--- will be lost in the single-particle picture discussed below (see Sect.~\ref{ss-SPD}).

Let us finally focus on the steady-state solution of the quantum-transport equation in (\ref{LvN-eff_L-cqp}).
Contrary to the global (carrier + quasiparticle) equation in (\ref{LvN-eff_L}), the identity operator
$\hat{\cal I}$ is no more a solution. Indeed, the latter fulfills the double commutator but not the
commutator-anticommutator structure in (\ref{Gamma_c-4}). Moreover, as we shall see in Sect.~\ref{s-PCE},
the steady-state solution of our effective transport equation is in general non-diagonal.
We also stress that, as for the global (carrier + quasiparticle) description, in the small-coupling limit
it is possible to show that the steady-state reduced density matrix is again positive-definite (see App.~\ref{a-PDC}).

Also for the present reduced description we can consider the
semiclassical or Boltzmann limit; as described in
Sect.~\ref{ss-SL}, this corresponds to neglecting the nondiagonal
matrix elements of the reduced density matrix, i.e.,
%\begin{equation}\label{SL4}
$\rho^c_{\nalphaone\nalphatwo} = f^c_\nalphaone
\delta_{\nalphaone\nalphatwo} . $%\end{equation}
Within such
approximation scheme the quantum-tran\-sport equation in
(\ref{Gamma_c-3}) reduces to the following Boltzmann equation for
the carrier subsystem:
\begin{equation}\label{SL5}
{d f^c_\nalpha \over dt} =
\sum_\nalphap \left(
P^c_{\nalpha\nalphap} f^c_\nalphap - P^c_{\nalphap\nalpha} f^c_\nalpha
\right)\ ,
\end{equation}
where
\begin{widetext}
\begin{equation}\label{P-bis}
P^c_{\nalpha\nalphap} = \sum_{\qu \pm}
{2\pi \over \hbar}
\left(N_\qu + {1 \over 2} \pm {1 \over 2}\right)
|{H}^{\qu}_{\nalpha\nalphap}|^2
\delta\left(\epsilon_\nalpha-\epsilon_\nalphap \pm \epsilon_\qu \right)
\end{equation}
\end{widetext}
are the usual carrier-quasiparticle semiclassical scattering rates given by the well-known
Fermi's golden rule.
We stress that, contrary to the global (carrier + quasiparticle) description considered in
Sect.~\ref{ss-SL}, the scattering rates in (\ref{P-bis}) are not symmetric:
$P^c_{\nalpha\nalphap} \neq P^c_{\nalphap\nalpha}$; this is a direct fingerprint of the irreversible
nature of the transport problem induced by energy-relaxation/dephasing processes.

\subsection{Single-particle description}\label{ss-SPD}

Most of the electronic properties of interest in the analysis of
charge-transport phenomena in semiconductor nanostructures are
single-particle quantities, i.e., physical quantities ascribed to
the generic particle in our electronic subsystem, like carrier
drift velocity, mean kinetic energy, etc. In this case, the
corresponding quantum-mechanical operator $\hat{A}$ is of the
form:
\begin{equation}\label{SP1}
\hat{A} = \sum_{\alpha\alpha'} A^{sp}_{\alpha\alpha'} \hat c^\dagger_{\alpha} \hat c^{ }_{\alpha'} \ .
\end{equation}
and its average value can be written as
%Inserting the above
%single-particle operator into Eq.~(\ref{average2-bis}) we get:
\begin{equation}\label{SP2}
A = \sum_{\alpha\alpha'} A^{sp}_{\alpha\alpha'} \rmtr\left\{\hat
c^\dagger_{\alpha} \hat c^{ }_{\alpha'} {\hat \rho} \right\} =
\sum_{\alpha\alpha'} A^{sp}_{\alpha\alpha'}
\rho^{sp}_{\alpha'\alpha} \ ,
\end{equation}
where
\begin{equation}\label{SP4}
\rho^{sp}_{\alpha_1\alpha_2} = \rmtr\left\{\hat
c^\dagger_{\alpha_2} \hat c^{ }_{\alpha_1} \hat{\rho}\right\}
\end{equation}
is the so-called single-particle density matrix.\cite{RMP} As we
can see, this is defined as average of the product of creation and
destruction operators; its diagonal elements ($\alpha_1 =
\alpha_2$) correspond to the single-particle carrier distribution
of the semiclassical Boltzmann theory, while the non diagonal
contributions ($\alpha_1 \neq \alpha_2$) describe
quantum-mechanical phase coherence between the single-particle
states $\alpha_1$ and $\alpha_2$. We stress that, while the
reduced density-matrix operator $\hat{\rho}^c$ describes the whole
many-electron system, the single-particle operator
$\hat{\rho}^{sp}$ provides an average or mean-field treatment of
the carrier subsystem; indeed the latter fails in describing
many-particle correlations, like Coulomb-correlation effects in
quasi zero-dimensional systems.\cite{Biolatti}

Since $\rho^{sp}_{\alpha_1\alpha_2}$ is the only quantity entering
the evaluation of the average value in (\ref{SP2}), it is
desirable to derive a corresponding equation of motion for the
single-particle density-matrix in (\ref{SP4}):
\begin{equation}\label{SP6}
{d \over dt} \rho^{sp}_{\alpha_1\alpha_2} = \rmtr\left\{\hat
c^\dagger_{\alpha_2} \hat c^{ }_{\alpha_1} {d \over dt} \hat{\rho}
\right\} \ .
\end{equation}
Inserting into the above expression the equation of motion for the
global density-matrix operator $\hat{\rho}$ in (\ref{LvN-eff_L})
we get:
\begin{equation}\label{SP7}
{d \over dt} \rho^{sp}_{\alpha_1\alpha_2} =
{d \over dt} \rho^{sp}_{\alpha_1\alpha_2}\Bigl|_{\hat{H}_\circ} +
{d \over dt} \rho^{sp}_{\alpha_1\alpha_2}\Bigl|_{\hat{C}} +
{d \over dt} \rho^{sp}_{\alpha_1\alpha_2}\Bigl|_{\hat{H}'} \ ,
\end{equation}
where
\begin{equation}\label{SP8}
{d \over dt} \rho^{sp}_{\alpha_1\alpha_2}\Bigl|_{\hat{H}_\circ} =
{1 \over i\hbar} \rmtr\left\{ \hat c^\dagger_{\alpha_2} \hat c^{
}_{\alpha_1} [\hat{H}_\circ, \hat{\rho}] \right\}
\end{equation}
is the time variation induced by the noninteracting Hamiltonian $\hat{H}_\circ$,
\begin{equation}\label{SP9}
{d \over dt} \rho^{sp}_{\alpha_1\alpha_2}\Bigl|_{\hat{C}} = -i\,
\rmtr\left\{ \hat c^\dagger_{\alpha_2} \hat c^{ }_{\alpha_1}
[\hat{\cal H}, \hat{\rho}^i(t)] \right\}
\end{equation}
is the contribution due to the quantum-correlation operator
$\hat{C}$ in (\ref{C-bis}), and
\begin{equation}\label{SP10}
{d \over dt} \rho^{sp}_{\alpha_1\alpha_2}\Bigl|_{\hat{H}'} = -
\rmtr\left\{ \hat c^\dagger_{\alpha_2} \hat c^{ }_{\alpha_1}
[\hat{\cal H},[\hat{\cal K}, \hat{\rho}]] \right\}
\end{equation}
is the time evolution dictated by the scattering superoperator $\Gamma$.

For a better evaluation of the various contributions in
(\ref{SP8})-(\ref{SP10}) it is convenient to expand the
commutators entering the trace, regrouping the various terms in a
different way. More specifically, by inserting the explicit form
of the single and double commutators, and using the cyclic
property of the trace, we finally get
%\begin{widetext}
\begin{equation}\label{SP11}
{d \over dt} \rho^{sp}_{\alpha_1\alpha_2}\Bigl|_{\hat{H}_\circ} =
{1 \over i\hbar} \rmtr\left\{ \left[\hat c^\dagger_{\alpha_2} \hat
c^{ }_{\alpha_1}, \hat{H}_\circ \right] \hat{\rho} \right\}\ ,
\end{equation}
\begin{equation}\label{SP12}
{d \over dt} \rho^{sp}_{\alpha_1\alpha_2}\Bigl|_{\hat{C}} = -i\,
\rmtr\left\{
 \left[ \hat c^\dagger_{\alpha_2} \hat c^{ }_{\alpha_1}, \hat{\cal H}\right] \hat{\rho}^i(t)
\right\} \ ,
\end{equation}
%\end{widetext}
and
\begin{equation}\label{SP13}
{d \over dt} \rho^{sp}_{\alpha_1\alpha_2}\Bigl|_{\hat{H}'} = -
\rmtr\left\{ \left[\left[\hat c^\dagger_{\alpha_2} \hat c^{
}_{\alpha_1}, \hat{\cal H}\right], \hat{\cal K}\right] \hat{\rho}
\right\} \ .
\end{equation}
As we can see, the various contributions to the time evolution of the single-particle density-matrix
can be written as global average values %[see Eq.~(\ref{average2-bis})]
of single as well as double-commutators.

The term in (\ref{SP11}) %---describing the time evolution dictated by the noninteracting Hamiltonian $\hat{H}_\circ$---
can be evaluated exactly. Indeed, recalling the explicit form of
the free-carrier + free-phonon Hamiltonian in (\ref{H0}) and using
the Fermionic anticommutation relations $\left\{\hat c^{ }_\alpha,
\hat c^\dagger_{\alpha'} \right\} = \delta_{\alpha\alpha'}$, we
obtain
\begin{equation}\label{SP14}
{d \over dt} \rho^{sp}_{\alpha_1\alpha_2}\Bigl|_{\hat{H}_\circ} =
{\epsilon_{\alpha_1}-\epsilon_{\alpha_2} \over i\hbar} \rho^{sp}_{\alpha_1\alpha_2} \ .
\end{equation}

In contrast, for the first- and second-order interaction
contributions in (\ref{SP12}) and (\ref{SP13}) it is not possible
to obtain closed equations of motion for the single-particle
density matrix $\rho^{sp}$: indeed such contributions involve
higher-order correlations, e.g., two-body and/or phonon-assisted
density matrices.\cite{RMP}  In order to get a closed equation for
$\rho^{sp}_{\alpha_1\alpha_2}$, an additional approximation is
needed, the so-called mean-field approximation. The latter
consists of a factorization of the higher-order correlation
functions into products of single-particle density matrices
$\rho^{sp}$ and/or quasiparticle populations $N_\qu$. The required
mean-field procedure and the explicit form of the resulting closed
equation of motion depend on the particular form of the
interaction Hamiltonian considered, e.g., carrier-carrier,
carrier-quasiparticle, etc. However, the free-evolution term
(\ref{SP11}) together with the first-order contribution in
(\ref{SP12}) describe, in general, coherent phenomena ---including
Hartree-Fock renormalization and coherent phonons--- while the
second-order term in (\ref{SP13}) describe
energy-relaxation/dephasing processes within the Markov
approximation.

At this point few comments are in order. The single-particle
description discussed in this section is based on the
Schr\"odinger picture: the equation of motion for
$\rho^{sp}_{\alpha_1\alpha_2}$ [see Eq.~(\ref{SP6})] is derived by
treating the operators $\hat c^\dagger$ and $\hat c$ in
(\ref{SP4}) as time-independent, while the time variation is fully
attributed to the density-matrix operator $\hat{\rho}$. Actually,
the most popular and commonly used approach~\cite{RMP} to derive
the equations of motion governing the time evolution of the
single-particle density matrix is based on the Heisenberg picture:
the density-matrix operator $\hat{\rho}$ entering Eq.~(\ref{SP4})
in the Heisenberg scheme is time-independent, while the time
evolution is fully ascribed to the Fermionic operators via their
corresponding Heisenberg equations of motion:
\begin{equation}\label{Heisenberg1}
{d \over dt} \hat c_\alpha = {1 \over i\hbar} [\hat c_\alpha, \hat{H}] \ .
\end{equation}
More precisely, within the Heisenberg picture Eq.~(\ref{SP6}) is replaced by:
\begin{eqnarray}\label{Heisenberg2}
{d \over dt} \rho^{sp}_{\alpha_1\alpha_2} = \rmtr\left\{ {d \over
dt} \left(\hat c^\dagger_{\alpha_2} \hat c^{ }_{\alpha_1}\right)
\hat{\rho} \right\} = {1 \over i\hbar} \rmtr\left\{ [\hat
c^\dagger_{\alpha_2} \hat c^{ }_{\alpha_1}, \hat{H}] \hat{\rho}
\right\} \ .
\end{eqnarray}
Contrary to the theoretical approach proposed in this paper, the
usual Heisenberg treatment is based on a correlation expansion of
the trace in (\ref{Heisenberg2}):\cite{RMP} starting again from
the Hamiltonian separation in (\ref{H}), a hierarchy of kinetic
equations involving higher-order density as well as
quasiparticle-assisted density matrices is established; the
different contributions are classified in terms of their
perturbation order. Such infinite hierarchy is truncated/closed
via the mean-field approximation previously recalled, and only at
this level the Markov limit is usually introduced.\cite{RMP} Aim
of this paper, in contrast, is to analyze the Markov limit from a
more general point of view; it is for this reason that the latter
has been introduced in very general terms in Sect.~\ref{ss-AMA}
before addressing any reduced description.

\subsubsection{Carrier-carrier interaction}\label{sss-CC-SP}

As first interaction mechanism we shall consider two-body Coulomb coupling.
The corresponding interaction Hamiltonian can be written as
\begin{equation}\label{calH_cc}
\hat{\cal H}^{cc} =
{1 \over 2 \hbar} \sum_{\alpha_1\alpha_2,\alpha'_1\alpha'_2} V^{cc}_{\alpha_1\alpha_2,\alpha'_1\alpha'_2}
\hat c^\dagger_{\alpha_1} \hat c^\dagger_{\alpha_2} \hat c^{ }_{\alpha'_1} \hat c^{ }_{\alpha'_2} \ ,
\end{equation}
where
$V^{cc}_{\alpha_1\alpha_2,\alpha'_1\alpha'_2}$
is the Coulomb matrix element for the generic two-body transition $\alpha'_1\alpha'_2 \to \alpha_1\alpha_2$.

In order to derive the explicit form of the second-order contribution to the single-particle dynamics in
(\ref{SP13}), two key quantities need to be evaluated:
the inner commutator
$\left[\hat c^\dagger_{\alpha_2} \hat c^{ }_{\alpha_1}, \hat{\cal H}\right]$
and the explicit form of the operator $\hat{\cal K}$.

More specifically, by employing the anticommutation relations for the Fermionic operators, we get:
\begin{widetext}
\begin{eqnarray}\label{inner_cc}
&&\left[ \hat c^\dagger_{\alpha_2} \hat c^{ }_{\alpha_1},
\hat{\cal H}^{cc}\right] = { 1 \over \hbar}
\sum_{\alpha_3\alpha_4\alpha_5}
\left(V^{cc}_{\alpha_1\alpha_3,\alpha_4\alpha_5} \hat
c^\dagger_{\alpha_2} \hat c^\dagger_{\alpha_3} \hat c^{
}_{\alpha_4} \hat c^{ }_{\alpha_5} \right. - \left.
V^{cc}_{\alpha_5\alpha_4,\alpha_3\alpha_2} \hat
c^\dagger_{\alpha_5} \hat c^\dagger_{\alpha_4} \hat c^{
}_{\alpha_3} \hat c^{ }_{\alpha_1} \right) \ .
\end{eqnarray}
\end{widetext}
The above commutator has the same structure of the interaction
Hamiltonian in (\ref{calH_cc}), i.e., it consists of a sum of
products of four Fermionic operators.

By inserting into Eq.~(\ref{calK3}) the explicit form of the two-body Coulomb interaction Hamiltonian in
(\ref{calH_cc}), we get:
\begin{widetext}
\begin{eqnarray}\label{calK_cc-1}
\hat{\cal K}^{cc} &=& {1 \over \hbar} \int_0^{t-t_0} d\tau {\hat
U}^{ }_\circ(\tau) \left( {1 \over 2}
\sum_{\alpha_1\alpha_2,\alpha'_1\alpha'_2}
V^{cc}_{\alpha_1\alpha_2,\alpha'_1\alpha'_2} \hat
c^\dagger_{\alpha_1} \hat c^\dagger_{\alpha_2} \hat c^{
}_{\alpha'_1} \hat c^{ }_{\alpha'_2} \right) {\hat
U}^\dagger_\circ(\tau) \ .
\end{eqnarray}
\end{widetext}
Recalling that
\begin{widetext}
\begin{eqnarray}\label{calK_cc-2}
&&{\hat U}^{ }_\circ(\tau) \hat c^\dagger_{\alpha_1} \hat
c^\dagger_{\alpha_2} \hat c^{ }_{\alpha'_1} \hat c^{ }_{\alpha'_2}
{\hat U}^\dagger_\circ(\tau) = \hat c^\dagger_{\alpha_1} \hat
c^\dagger_{\alpha_2} \hat c^{ }_{\alpha'_1} \hat c^{ }_{\alpha'_2}
e^{(\epsilon_{\alpha_1}+\epsilon_{\alpha_2}-\epsilon_{\alpha'_1}-\epsilon_{\alpha'_2})
\tau \over i \hbar} \ ,
\end{eqnarray}
\end{widetext}
we finally obtain
\begin{equation}\label{calK_cc-3}
\hat{\cal K}^{cc} = {2\pi \over \hbar}
\sum_{\alpha_1\alpha_2,\alpha'_1\alpha'_2}
{1 \over 2} V^{cc}_{\alpha_1\alpha_2,\alpha'_1\alpha'_2}
{\cal D}^{cc}_{\alpha_1\alpha_2,\alpha'_1\alpha'_2}
\hat c^\dagger_{\alpha_1} \hat c^\dagger_{\alpha_2} \hat c^{ }_{\alpha'_1} \hat c^{ }_{\alpha'_2}
\end{equation}
with
\begin{equation}\label{calD_cc}
{\cal D}^{cc}_{\alpha_1\alpha_2,\alpha'_1\alpha'_2} = {1 \over 2\pi} \int_0^{t-t_0}
e^{(\epsilon_{\alpha_1}+\epsilon_{\alpha_2}-\epsilon_{\alpha'_1}-\epsilon_{\alpha'_2}) \tau \over i \hbar} d\tau \ .
\end{equation}
We get again the same operatorial structure: a sum of products of four Fermionic operators.

Given the two results in (\ref{inner_cc}) and (\ref{calK_cc-3}), their commutator ---key ingredient in
Eq.~(\ref{SP13})--- will involve, in general, products of six Fermionic
operators.

As anticipated, in order to get a closed equation of motion for
the single-particle density matrix $\rho^{sp}$ we are forced to
employ the mean-field approximation; the latter allows in this
case to write, in general, the average values of six Fermionic
operators as products of three single-particle density-matrix
elements.

By applying such mean-field factorization procedure to the
explicit form of the outer com\-mu\-ta\-tor in (\ref{SP13}), the
final result ---not reported here--- can be cast into the general
form:
\begin{equation}\label{eom_cc}
{d \over dt} \rho^{sp}_{\alpha_1\alpha_2}\Bigl|_{cc} = {\cal
F}^{cc,{\rm in}}[\rho^{sp}]_{\alpha_1\alpha_2}  - {\cal
F}^{cc,{\rm out}}[\rho^{sp}]_{\alpha_1\alpha_2} \,\, .
\end{equation}
As for the case of the reduced description (see
Sect.~\ref{ss-RD}), the time variation of the single-particle
density matrix is the sum of a positive ---in-scattering--- and a
negative ---out-scattering--- contribution. However, contrary to
the global and reduced descriptions previously considered, now the
in- and out-scattering contributions are non-linear functions of
the single-particle density matrix $\rho^{sp}$. In particular, in
this case of two-body interaction between a main (M) and a partner
(P) carrier, the superoperators ${\cal F}^{cc,{\rm in}}$ and
${\cal F}^{cc,{\rm out}}$ both involve a product structure of the
form $\hat{\rho}^{sp_{\rm M}} \hat{\rho}^{sp_{\rm P}} (\hat{\cal
I}-\hat{\rho}^{sp_{\rm M}})(\hat{\cal I}-\hat{\rho}^{sp_{\rm
P}})$.

Also for the present single-particle description it is possible to
consider the semiclassical or Boltzmann limit introduced in
Sect.~\ref{ss-SL}. This amounts again to neglect non-diagonal
density-matrix elements: $\rho^{sp}_{\alpha_1\alpha_2} =
f^{sp}_{\alpha_1} \delta_{\alpha_1\alpha_2} $ . By inserting the
above diagonal form of $\rho^{sp}$ into the in- and out-scattering
superoperators ${\cal F}^{cc,{\rm in}}$ and ${\cal F}^{cc,{\rm
out}}$, the following Boltzmann-like equation for the
semiclassical single-particle distribution $f^{sp}_\alpha$ may be
derived:
\begin{equation}\label{SL_cc}
{d f^{sp}_\alpha \over dt} =
\sum_{\alpha'} \left[
(1-f^{sp}_\alpha)
P^{cc}_{\alpha\alpha'} f^{sp}_{\alpha'} -
(1-f^{sp}_{\alpha'})
P^{cc}_{\alpha'\alpha} f^{sp}_\alpha
\right]\ ,
\end{equation}
where
\begin{widetext}
\begin{equation}\label{P_cc}
P^{cc}_{\alpha\alpha'} = {2\pi \over \hbar}
\sum_{\overline{\alpha} \, \overline{\alpha}'}
(1-f^{sp}_{\overline{\alpha}}) \left|
V^{cc}_{\alpha\overline{\alpha},\alpha'\overline{\alpha}'}
\right|^2 f^{sp}_{\overline{\alpha}'}
\delta(\epsilon_\alpha+\epsilon_{\overline{\alpha}} -
\epsilon_{\alpha'} -\epsilon_{\overline{\alpha}'})
\end{equation}
\end{widetext}
are two-body carrier-carrier scattering rates describing the main-carrier transition $\alpha' \to \alpha$
accompanied by the partner-carrier transition $\overline{\alpha}' \to \overline{\alpha}$.
As we can see, also in the semiclassical limit we deal with a non-linear transport equation; such
nonlinearities are ascribed (i) to the presence of the carrier distribution $f^{sp}$ of the initial
partner carrier, and (ii) to the two Pauli-blocking factors
$(1-f^{sp})$
corresponding to the final states of both main and partner carriers.
Comparing the semiclassical transport equation in (\ref{SL_cc}) to its quantum-mechanical generalization in
(\ref{eom_cc}), we clearly see that the various terms of the form
$(\delta_{\alpha\alpha'} - \rho^{sp}_{\alpha\alpha'})$
are the natural generalization of the Pauli-blocking factors $(1-f^{sp}_\alpha)$ of the semiclassical theory.

\subsubsection{Carrier-quasiparticle interaction}\label{sss-CQP-SP}

Let us now come to the carrier-quasiparticle coupling mechanism. By adopting as explicit form of the
carrier-quasiparticle quantity
$\hat{\cal H}_\qu = \hat{\cal H}^\dagger_{-\qu}$ the single-particle operator
\begin{equation}\label{calH-qu}
\hat{\cal H}_\qu = { 1 \over \hbar} \sum_{\alpha\alpha'}
g_{\alpha\alpha',\qu} \hat c^\dagger_{\alpha} \hat c^{ }_{\alpha'} \ ,
\end{equation}
the carrier-quasiparticle interaction Hamiltonian in (\ref{H_c-qp}) is given by:
\begin{equation}\label{calH_cqp}
\hbar \hat{\cal H}^{c-qp} = \sum_{\alpha\alpha',\qu}
\left(
g_{\alpha\alpha',\qu}
\hat c^\dagger_{\alpha} \hat b^{ }_\qu \hat c^{ }_{\alpha'} +
g^*_{\alpha\alpha',\qu} \hat c^\dagger_{\alpha'}\hat b^\dagger_\qu \hat c^{ }_{\alpha} \right) \ ,
\end{equation}
where
\begin{equation}\label{gqu}
g_{\alpha\alpha',\qu} = \overline{g}_\qu f_{\alpha\alpha',\qu} = \overline{g}_\qu \int \phi^*_\alpha({\bf r})
e^{i \qu \cdot {\bf r}} \phi_{\alpha'}({\bf r}) d{\bf r}
\end{equation}
is the carrier-quasiparticle matrix element for the
single-particle transition $\alpha' \to \alpha$ induced by the
quasiparticle bulk mode $\qu$. The explicit form of the coupling
function $\overline{g}$ depends on the particular
carrier-quasiparticle interaction mechanism considered. In any
case we have: $g_{\alpha\alpha',\qu} = g^*_{\alpha'\alpha, -\qu}$.
As for carrier-carrier interaction, in order to derive the
second-order contribution to the single-particle dynamics in
(\ref{SP13}), we shall evaluate the inner commutator $\left[\hat
c^\dagger_{\alpha_2} \hat c^{ }_{\alpha_1}, \hat{\cal H}\right]$
as well as the operator $\hat{\cal K}$.

More specifically, by employing again the anticommutation relations for the Fermionic operators we get:
\begin{widetext}
\begin{eqnarray}\label{inner_cqp}
\left[ \hat c^\dagger_{\alpha_2} \hat c^{ }_{\alpha_1}, \hat{\cal
H}^{c-qp}\right] = -{1 \over \hbar} \sum_{\alpha_3,\qu} \left(
g_{\alpha_3\alpha_2,\qu} \hat c^\dagger_{\alpha_3} \hat b^{ }_\qu
\hat c^{ }_{\alpha_1} + g^*_{\alpha_2\alpha_3,\qu} \hat
c^\dagger_{\alpha_3} \hat b^\dagger_\qu \hat c^{ }_{\alpha_1}
-g_{\alpha_1\alpha_3,\qu} \hat c^\dagger_{\alpha_2} \hat b^{ }_\qu
\hat c^{ }_{\alpha_3} - g^*_{\alpha_3\alpha_1,\qu} \hat
c^\dagger_{\alpha_2} \hat b^\dagger_\qu \hat c^{ }_{\alpha_3}
\right)\ .
\end{eqnarray}
\end{widetext}
As we can see, the commutator has the same operatorial structure of the carrier-quasiparticle Hamiltonian, i.e.,
it involves products of one Bosonic and two Fermionic operators.

By inserting into Eq.~(\ref{calK3}) the explicit form of the
carrier-quasiparticle interaction Hamiltonian in (\ref{calH_cqp}),
we get:
\begin{widetext}
\begin{eqnarray}\label{calK_cqp-1}
\hat{\cal K}^{c-qp} &=& {1 \over \hbar} \int_0^{t-t_0} d\tau {\hat
U}^{ }_\circ(\tau) \left(\sum_{\alpha\alpha',\qu}
g_{\alpha\alpha',\qu} \hat c^\dagger_{\alpha} \hat b^{ }_\qu \hat
c^{ }_{\alpha'} \right) {\hat U}^\dagger_\circ(\tau) + \mbox{h.c.}
\ ,
\end{eqnarray}
\end{widetext}
where h.c. denotes its hermitian conjugate. Recalling that
\begin{equation}\label{calK_cqp-2}
{\hat U}^{ }_\circ(\tau)
\hat c^\dagger_{\alpha} \hat b_\qu \hat c^{ }_{\alpha'}
{\hat U}^\dagger_\circ(\tau) =
\hat c^\dagger_{\alpha} \hat b_\qu \hat c^{ }_{\alpha'}
e^{(\epsilon_{\alpha}-\epsilon_{\alpha'} -\epsilon_\qu) \tau \over i \hbar}
\ ,
\end{equation}
we finally obtain
\begin{equation}\label{calK_cqp-3}
\hat{\cal K}^{c-qp} = {2\pi \over \hbar}
\sum_{\alpha\alpha',\qu}
g_{\alpha\alpha',\qu}
{\cal D}^{c-qp, -}_{\alpha\alpha',\qu}
\hat c^\dagger_{\alpha} \hat b_\qu \hat c^{ }_{\alpha'}
+ \mbox{h.c.} \ ,
\end{equation}
with
\begin{equation}\label{calD_cqp}
{\cal D}^{c-qp, \pm}_{\alpha\alpha',\qu} = {1 \over 2\pi} \int_0^{t-t_0}
e^{(\epsilon_{\alpha}-\epsilon_{\alpha'} \pm\epsilon_\qu) \tau \over i \hbar} d\tau \ .
\end{equation}
We get again the same operatorial structure: one Bosonic times two Fermionic operators.

Given the two results in (\ref{inner_cqp}) and (\ref{calK_cqp-3}),
their commutator ---key ingredient in Eq.~(\ref{SP13})--- will involve products of two Bosonic and two
Fermionic operators as well as products of four Fermionic operators.
As for the carrier-carrier interaction previously discussed, in order to get a closed equation of
motion for the single-particle density matrix $\rho^{sp}$ we are forced to employ again the mean-field
approximation; the latter allows to write (i) the average value of two Fermionic times two Bosonic operators
as the product of single-particle density-matrix elements $\rho^{sp}_{\alpha\alpha'}$ times quasiparticle
distributions, and (ii) the average values of four Fermionic operators as products of two single-particle
density-matrix elements.

By applying such mean-field factorization procedure to the
explicit form of the outer commutator in (\ref{SP13}), the final
result ---not reported here--- can be cast in the same form of the
one for carrier-carrier interaction in (\ref{eom_cc}), i.e.,
\begin{equation}\label{eom_cqp}
{d \over dt} \rho^{sp}_{\alpha_1\alpha_2}\Bigl|_{c-qp} = {\cal
F}^{c-qp,{\rm in}}[\rho^{sp}]_{\alpha_1\alpha_2}  - {\cal
F}^{c-qp,{\rm out}}[\rho^{sp}]_{\alpha_1\alpha_2} \,\, .
\end{equation}
As for the case of carrier-carrier interaction, the above in- and
out-scattering contributions are again non-linear functions of the
single-particle density matrix $\rho^{sp}$. More specifically,
their general structure is of the form $\hat{\rho}^{sp}(\hat{\cal
I}-\hat{\rho}^{sp})$; Such nonlinearities
---ascribed to Pauli-blocking effects--- vanish in the so-called
low-density limit ($\hat{\cal I}-\hat{\rho}^{sp} \rightarrow
\hat{\cal I}$).

In the semiclassical limit, by inserting the diagonal
density-matrix form into the in- and out- scattering functionals
${\cal F}^{c-qp,{\rm in}}$ and ${\cal F}^{c-qp,{\rm out}}$, we get
a non-linear Boltzmann equation formally identical to the
single-particle transport equation in (\ref{SL_cc}):
\begin{equation}\label{SL_cqp}
{d f^{sp}_\alpha \over dt} =
\sum_{\alpha'} \left[
(1-f^{sp}_\alpha)
P^{c-qp}_{\alpha\alpha'} f^{sp}_{\alpha'} -
(1-f^{sp}_{\alpha'})
P^{c-qp}_{\alpha'\alpha} f^{sp}_\alpha
\right]\ ,
\end{equation}
where
\begin{equation}\label{P_cqp}
P^{c-qp}_{\alpha\alpha'} = {2\pi \over \hbar} \sum_{\qu \pm}
\left|
g_{\alpha\alpha',\qu}
\right|^2
\delta(\epsilon_\alpha- \epsilon_{\alpha'} \pm \epsilon_\qu)
\end{equation}
denote semiclassical carrier-quasiparticle scattering rates for
the single-particle transition $\alpha' \to \alpha$. We deal again
with a non-linear transport equation; in this case, however, such
nonlinearities are only ascribed to the Pauli-blocking factor of
the final state, and they vanish in the low-density limit as
discussed in the following subsection.

\subsubsection{Low-density limit}\label{sss-LDL}

Let us finally consider the so-called ``low-density limit''. To this aim, let us recall that within the
single-particle description previously introduced the average occupation number for the generic state
$\alpha$ is simply given by the diagonal elements of the single particle density matrix in (\ref{SP4}):
\begin{equation}\label{AON}
\rho^{sp}_{\alpha\alpha} = \rmtr\left\{\hat c^\dagger_{\alpha}
\hat c^{ }_{\alpha} \hat{\rho}\right\} = f^{sp}_\alpha \ .
\end{equation}
It is then clear that at low carrier concentrations (low
densities) the magnitude of all density-matrix elements is much
smaller than 1; More precisely, we have:
$\rho^{sp}_{\alpha\alpha'} \to 0 \ , %\qquad
(\delta_{\alpha\alpha'} - \rho^{sp}_{\alpha\alpha'}) \to
\delta_{\alpha\alpha'} \ .$

In this limit the carrier-carrier scattering contributions in
%(\ref{calF_cc-in}) and (\ref{calF_cc-out})
(\ref{eom_cc}) vanish since, as anticipated, they involve the
single-particle density matrix of the partner carrier
$\hat{\rho}^{sp_{\rm P}}$. As a result, in the low-density limit
the only non-vanishing contribution to the interaction dynamics is
given by the carrier-quasiparticle terms in (\ref{eom_cqp}). More
precisely, by inserting the low-density condition %(\ref{LDL1})
in our quantum transport equation (\ref{eom_cqp})
as well as in
the corresponding in- and out-scattering functions, we finally get the
following linear transport equation:
\begin{widetext}
\begin{equation}\label{LDL-eom}
{d \over dt} \rho^{sp}_{\alpha_1\alpha_2}\Bigl|_{c-qp} =
\sum_{\alpha'_1\alpha'_2} \Gamma_{\alpha_1\alpha_2,\alpha'_1\alpha'_2} \rho^{sp}_{\alpha'_1\alpha'_2} =
\sum_{\alpha'_1\alpha'_2}
\Gamma^{\rm in}_{\alpha_1\alpha_2,\alpha'_1\alpha'_2} \rho^{sp}_{\alpha'_1\alpha'_2}
- \sum_{\alpha'_1\alpha'_2} \Gamma^{\rm out}_{\alpha_1\alpha_2,\alpha'_1\alpha'_2}
\rho^{sp}_{\alpha'_1\alpha'_2}
\end{equation}
with
\begin{equation}\label{LDL-in}
\Gamma^{\rm in}_{\alpha_1\alpha_2,\alpha'_1\alpha'_2} = {2\pi \over \hbar^2}
\sum_{\qu \pm}
\left(N_\qu +{1 \over 2} \pm {1 \over 2}\right)
\left(
g_{\alpha_1\alpha'_1,\qu}
g^*_{\alpha_2\alpha'_2,\qu}
{\cal D}^{c-qp,\pm *}_{\alpha_2\alpha'_2,\qu}
+
 {\cal D}^{c-qp,\pm}_{\alpha_1\alpha'_1,\qu}
g_{\alpha_1\alpha'_1,\qu}
g^*_{\alpha_2\alpha'_2,\qu}
\right)
\end{equation}
and
\begin{equation}\label{LDL-out}
\Gamma^{\rm out}_{\alpha_1\alpha_2,\alpha'_1\alpha'_2} \!=\! {2\pi
\over \hbar^2} \!\sum_{\alpha'', \qu \pm} \!\left(N_\qu +{1 \over
2} \pm {1 \over 2}\right)\! \left( g^*_{\alpha''\alpha_1,\qu}
g_{\alpha''\alpha'_1,\qu} {\cal
D}^{c-qp,\pm}_{\alpha''\alpha'_1,\qu} \!\delta_{\alpha_2\alpha'_2}
\!+\! \delta_{\alpha_1\alpha'_1}\! {\cal D}^{c-qp,\pm
*}_{\alpha''\alpha'_2,\qu} g^*_{\alpha''\alpha'_2,\qu}
g_{\alpha''\alpha_2,\qu} \right) \ .
\end{equation}
\end{widetext}
We stress that the above linear in- and out-scattering operators are formally identical to the
reduced-description ones in Eqs.~(\ref{in2_c}) and (\ref{out2_c}).
This can be easily understood, considering that the present low-density limit is physically equivalent
to consider a system of just one electron interacting with the quasiparticle degrees of freedom;
this would correspond to substitute the many-electron configuration label $\nalpha$ with the state
$\alpha$ of the only electron considered.

Combining Eqs.~(\ref{SP14}) and (\ref{LDL-eom}) and neglecting the first-order term in (\ref{SP12}),
we finally get the following quantum-transport equation:
\begin{equation}\label{LDL-final}
{d \over dt} \rho^{sp}_{\alpha_1\alpha_2} = {\cal L}_{\alpha_1\alpha_2,\alpha'_1\alpha'_2}
\rho^{sp}_{\alpha'_1\alpha'_2} \ ,
\end{equation}
where
\begin{equation}\label{calL_sp}
{\cal L}_{\alpha_1\alpha_2,\alpha'_1\alpha'_2} = {\epsilon_{\alpha_1}-\epsilon_{\alpha_2}
\over i\hbar} \delta_{\alpha_1\alpha_2,\alpha'_1\alpha'_2} + \Gamma_{\alpha_1\alpha_2,\alpha'_1\alpha'_2}
\end{equation}
is the effective single-particle Liouville operator in the  low-density limit.

\section{Analysis of phase-coherence effects in semiconductor nanostructures}\label{s-PCE}

Let us now discuss in more detail the physical meaning of the in- and out-scattering operators
in Eqs.~(\ref{LDL-in})-(\ref{LDL-out}).
As shown in Eq.~(\ref{LDL-eom}), the generic matrix element
$\Gamma^{\rm in/out}_{\alpha_1\alpha_2,\alpha_1'\alpha_2'}$ describes how the density-matrix
element $\rho_{\alpha_1'\alpha_2'}$ influences the time evolution of the element $\rho_{\alpha_1\alpha_2}$.
We shall try to classify such matrix elements according to their role and physical meaning.

As a first class of terms, let us consider the semiclassical ones ($\alpha_1\alpha_1' =\alpha_2\alpha_2'$):
\begin{equation}\label{T_1}
\gamma^{T_1}_{\alpha\alpha'} = \Gamma_{\alpha\alpha,\alpha'\alpha'} =
P_{\alpha\alpha'} - \delta_{\alpha\alpha'} \sum_{\alpha''} P_{\alpha''\alpha'}\ .
\end{equation}
As discussed in App.~\ref{a-PDC}, the latter, also referred to as
$T_1$ contributions, are fully expressed in terms of the
semiclassical rates $P_{\alpha\alpha'}$ [see Eq.~(\ref{P_cqp})]
and act on the carrier distribution $f_\alpha =
\rho^{sp}_{\alpha\alpha}$ only giving rise to electron and energy
redistribution within the traditional Boltzmann picture.

A second class of terms is given by the so-called $T_2$ contributions
($\alpha_1 \alpha_2 = \alpha'_1\alpha'_2$):
\begin{equation}\label{T_2}
\gamma^{T_2}_{\alpha_1\alpha_2} = \Gamma_{\alpha_1\alpha_2,\alpha_1\alpha_2} =
-\left(\sum_{\alpha''} P_{\alpha''\alpha_1} +\sum_{\alpha''} P_{\alpha''\alpha_2}\right) \ .
\end{equation}
In the absence of other $\Gamma$ matrix elements, these $T_2$
contributions describe the damping of non-diagonal density-matrix
elements $p_{\alpha_1\alpha_2} =
\rho^{sp}_{\alpha_1\neq\alpha_2}$, also called inter-state
polarizations;\cite{RMP} Indeed, from Eq.~(\ref{LDL-eom}) we
easily get:
\begin{equation}\label{T_2-eom}
{d \over dt} p_{\alpha_1\alpha_2}\Bigl|_{c-qp} = \gamma^{T_2}_{\alpha_1\alpha_2} p_{\alpha_1\alpha_2} \ ,
\end{equation}
whose solution is simply given by:
\begin{equation}\label{T_2-expsol}
p_{\alpha_1\alpha_2}(t) = e^{\gamma^{T_2}_{\alpha_1\alpha_2} (t-t_0)} p_{\alpha_1\alpha_2}(t_0) \ .
\end{equation}
The above exponential decay ($\gamma^{T_2}_{\alpha_1\alpha_2}$ is
always negative) of non-diagonal density-matrix elements or
inter-state polarizations is the well-known decoherence or
dephasing process.\cite{RMP}
We stress that, in spite of
the quantum-mechanical (i.e. coherent) character of
$p_{\alpha_1\alpha_2}$, the dephasing rates $\gamma^{T_2}$ in
(\ref{T_2}) involve semiclassical ingredients only; indeed, the
latter can be regarded as the sum of the two semiclassical
out-scattering rates for states $\alpha_1$ and $\alpha_2$ (see
App.~\ref{a-PDC}).

In addition to these two classes of terms, acting independently on $f_\alpha$ and
$p_{\alpha_1\alpha_2}$, the fully non-diagonal description in (\ref{LDL-eom}) involves also non
trivial coupling contributions, namely
(i) coupling between different polarizations ($p_{\alpha_1\alpha_2} \to p_{\alpha'_1\alpha'_2}$),
and (ii)
terms of the form $\alpha\alpha \to \alpha_1\alpha_2$ and vice versa; the latter describe coupling
between $f_\alpha$ and
$p_{\alpha_1\alpha_2}$.
Denoting with $T_3$-contributions these last $f \to p$ terms and neglecting $p \to p'$ couplings,
the quantum-transport equation in (\ref{LDL-eom}) can be schematically written as:
\begin{equation}\label{eom-schem}
{d \over dt} \left(\matrix{f \cr p}\right) =
\left(
\matrix{\gamma^{T_1} & \gamma^{T_3} \cr
\gamma^{T_3} & \gamma^{T_2}}
\right)
\left(\matrix{f \cr p}\right) \ ,
\end{equation}
where $\gamma^{T_3}$ denotes schematically all the $f \to p$ coupling terms previously mentioned.

We clearly see that, in the absence of $T_3$ terms ($\gamma^{T_3}
= 0$) there is absolutely no coupling between population ($f$) and
polarization ($p$). This approximation scheme, known as $T_1 T_2$
model,\cite{RMP} is the most popular model used for the
interpretation of ultrafast optical experiments in semiconductors.

In contrast, the introduction of these $T_3$ terms ---not included
in the conventional $T_1 T_2$ model--- gives rise to a non-trivial
coupling between $f$ and $p$, which manifests itself in a residual
single-particle phase coherence also in steady-state conditions.
More specifically, while within the $T_1 T_2$ model in the limit
$t \to \infty$ all polarization terms $p_{\alpha_1\alpha_2}$
vanish [see Eq.~(\ref{T_2-expsol})] and the resulting steady-state
density matrix is diagonal ($\rho^{sp}_{\alpha_1\alpha_2} =
f_{\alpha_1} \delta_{\alpha_1\alpha_2}$), in the presence of $T_3$
terms we may have a steady-state solution characterized by
non-zero values of $p_{\alpha_1\alpha_2}$ (see below). As
discussed in App.~\ref{a-PDC}, in the small-coupling limit the
steady-state solution $\rho^{sp}_{\alpha_1\alpha_2}$ of the
transport equation in (\ref{LDL-final}) is always
positive-definite. This suggests the introduction of the unitary
transformation $T_{\alpha\bar{\alpha}} = \langle \alpha\vert
\bar{\alpha} \rangle$ which diagonalizes the steady-state solution
$\rho^{sp}$. This allows to clearly express the positive character
of the solution according to Eq.~(\ref{PD}).

In order to gain more insight into the fully non-diagonal density-matrix approach presented so far,
let us consider an extremely simplified scenario:
an electronic two-level system interacting with a single phonon mode $\qu$.
In this case we deal with a two-by-two density matrix of the form:
\begin{equation}\label{rho2D}
\rho^{sp} =
\left(\matrix{
\rho^{sp}_{aa} & \rho^{sp}_{ab} \cr
\rho^{sp}_{ba} & \rho^{sp}_{bb}
}\right)
=
\left(\matrix{
f_a & p^* \cr
p^{ } & f_b
}\right) \ .
\end{equation}
Here, the diagonal elements $\rho^{sp}_{aa}$ and $\rho^{sp}_{bb}$ coincide with the semiclassical
ground- and excited-state level populations $f_a$ and $f_b$, while the non diagonal element
$p = \rho^{sp}_{ba}$ (together with its complex conjugate $p^* = \rho^{sp}_{ab}$) describes the
degree of quantum-mechanical phase coherence between states/levels $a$ and $b$.
Let us introduce the following (arbitrary) $\nu = \{\alpha_1,\alpha_2\}$ mapping:
$
1 = \{a,a\},
2 = \{b,b\},
3 = \{b,a\},
4 = \{a,b\}
$.
Within such representation, the two-by-two density matrix in (\ref{rho2D}) is mapped into a
four-dimensional vector, and the single-particle Liouville superoperator
${\cal L}$ in (\ref{calL_sp}) will correspond to a four-by-four matrix.
More specifically, within the four-dimensional mapping given above the transport equation (\ref{LDL-final})
in steady-state conditions reduces to the following homogeneous linear problem:
\begin{equation}\label{eom-4D}
\left( \matrix{ {\cal L}_{aa,aa} & {\cal L}_{aa,bb} & {\cal
L}_{aa,ba} & {\cal L}_{aa,ab} \cr {\cal L}_{bb,aa} & {\cal
L}_{bb,bb} & {\cal L}_{bb,ba} & {\cal L}_{bb,ab} \cr {\cal
L}_{ba,aa} & {\cal L}_{ba,bb} & {\cal L}_{ba,ba} & {\cal
L}_{ba,ab} \cr {\cal L}_{ab,aa} & {\cal L}_{ab,bb} & {\cal
L}_{ab,ba} & {\cal L}_{ab,ab} }\right) \left( \matrix{ f_a \cr f_b
\cr p \cr p^* }\right) = 0
\end{equation}
with
\begin{widetext}
\begin{equation}\label{calL-4D}
\left( \matrix{ {\cal L}_{aa,aa} & {\cal L}_{aa,bb} & {\cal
L}_{aa,ba} & {\cal L}_{aa,ab} \cr {\cal L}_{bb,aa} & {\cal
L}_{bb,bb} & {\cal L}_{bb,ba} & {\cal L}_{bb,ab} \cr {\cal
L}_{ba,aa} & {\cal L}_{ba,bb} & {\cal L}_{ba,ba} & {\cal
L}_{ba,ab} \cr {\cal L}_{ab,aa} & {\cal L}_{ab,bb} & {\cal
L}_{ab,ba} & {\cal L}_{ab,ab} }\right) = {1\over i\hbar} \left(
\matrix{ 0 & 0 & 0 & 0 \cr 0 & 0 & 0 & 0 \cr 0 & 0 &
\Delta\epsilon & 0 \cr 0 & 0 & 0 & -\Delta\epsilon }\right) +
\left( \matrix{ \Gamma_{aa,aa} & \Gamma_{aa,bb} & \Gamma_{aa,ba} &
\Gamma_{aa,ab} \cr \Gamma_{bb,aa} & \Gamma_{bb,bb} &
\Gamma_{bb,ba} & \Gamma_{bb,ab} \cr \Gamma_{ba,aa} &
\Gamma_{ba,bb} & \Gamma_{ba,ba} & \Gamma_{ba,ab} \cr
\Gamma_{ab,aa} & \Gamma_{ab,bb} & \Gamma_{ab,ba} & \Gamma_{ab,ab}
}\right)\ .
\end{equation}
\end{widetext}
Here $\Delta\epsilon = \epsilon_b-\epsilon_a$ denotes the interlevel energy splitting.
Starting from the explicit form of the elec\-tron-qua\-si\-par\-ti\-cle in- and
out-scat\-te\-ring su\-per\-ope\-ra\-tors in (\ref{LDL-in}-\ref{LDL-out}) and assuming
real and symmetric single-particle wavefunctions $\phi_\alpha({\bf r})$ in Eq.~(\ref{gqu}),
the resulting expression for the matrix elements of $\Gamma$ for our two-level system is of the form:
\begin{widetext}
\begin{equation}\label{Gamma-4D}
\left( \matrix{ \Gamma_{aa,aa} & \Gamma_{aa,bb} & \Gamma_{aa,ba} &
\Gamma_{aa,ab} \cr \Gamma_{bb,aa} & \Gamma_{bb,bb} &
\Gamma_{bb,ba} & \Gamma_{bb,ab} \cr \Gamma_{ba,aa} &
\Gamma_{ba,bb} & \Gamma_{ba,ba} & \Gamma_{ba,ab} \cr
\Gamma_{ab,aa} & \Gamma_{ab,bb} & \Gamma_{ab,ba} & \Gamma_{ab,ab}
}\right) = \frac{\epsilon^{cp}}{\hbar} \left( \matrix{ - N &
(N+1) & 0 & 0 \cr
 N & - (N+1) & 0 & 0 \cr \frac{i}{2} \xi N & -\frac{i}{2} \xi (N+1) &
-\frac{1}{2} (2N+1) & \frac{1}{2} (2N+1) \cr -\frac{i}{2} \xi N &
\frac{i}{2} \xi (N+1) & \frac{1}{2} (2N+1) & -\frac{1}{2} (2N+1)
}\right) \ ,
\end{equation}
\end{widetext}
where $\xi = \frac{g_{ab,\bar{\bf q}}(g_{aa,\bar{\bf
q}}-g_{bb,\bar{\bf q}})}{|g_{ab,\bar{\bf q}}|^2}$ (with
$\epsilon_{\bar{\qu}}= \Delta\epsilon$), $\epsilon^{cp}$ denotes
the carrier-phonon coupling constant, and $N$ is the Bose
occupation number in (\ref{B-H}) for our two-level system.

As already stressed, by neglecting $f \to p$ terms (which
corresponds to set $g_{aa,\bar{\bf q}}=g_{bb,\bar{\bf q}}$),
populations and polarizations are totally decoupled. In contrast,
in the presence of these $T_3$ terms our steady-state solution
exhibits a residual single-particle coherence, i.e., a
polarization $p$ different from zero.

\begin{figure}
\includegraphics[width=0.4\textwidth]{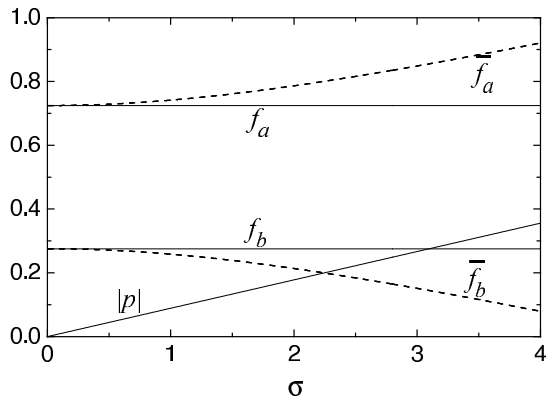}
\caption{\label{fig-tls} Phonon-induced single-particle phase
coherence for an isolated quantum-dot system: modulus of the
interlevel polarization $|p|$ (solid curve) and dressed-states
populations $\bar{f}_b$ and $\bar{f}_a$ (dashed curves) as a
function of the coupling-constant ratio $\sigma$, for an
interlevel energy splitting $\Delta \epsilon$ = 25 meV at room
temperature.}
\end{figure}
Figure \ref{fig-tls} shows the modulus of the interlevel
polarization $p$ as a function of the dimensionless coupling
parameter $\sigma = {\epsilon^{cp} \over \Delta\epsilon}$. In this
numerical example we have chosen $\Delta\epsilon = 25$\,meV, $T =
300$\,K, and $a = b = 0.25$. As we can see, $|p|$ comes out to be
proportional to the coupling parameter $\sigma$, as can be readily
verified by a closer inspection of our four-by-four superoperator
${\cal L}$ in (\ref{calL-4D}). It follows that for any finite
value of the coupling parameter $\sigma$, we deal with a
non-diagonal steady-state solution $\rho$ [see Eq.~(\ref{rho2D})].
More specifically, its diagonal elements $f_a$ and $f_b$ are
$\sigma$-independent and correspond to the semiclassical
equilibrium distribution: ${f_b \over f_a} = {N \over N+1}$; the
non-diagonal elements $p$ and $p^*$ will increase linearly with
the coupling parameter $\sigma$. In the small-coupling limit
($\sigma \ll 1$) such non-diagonal contributions are much smaller
than the diagonal ones. For increasing values of $\sigma$ we
progressively enter the strong-coupling regime, up to the point
where our two-by-two density matrix in (\ref{rho2D}) is no more
positive definite. Indeed, for $|p| > \sqrt{f_a f_b}$, the
determinant of $\rho$ (i.e., $f_a f_b-|p|^2$) becomes negative. As
anticipated, this is exactly the regime for which the Markov
approximation itself is no longer valid. However, for small and
moderate values of $\sigma$ our two-by-two density matrix is
positive-definite, which suggests the introduction of a
dressed-state basis in which the latter is diagonal. The new
populations $\overline{f}_a$ and $\overline{f}_b$ (dashed curves
in Fig.~\ref{fig-tls}) can be regarded as the average occupation
of such dressed states. As we can see, for $\sigma = 0$ they
coincide with the noninteracting thermal ones; for increasing
values of the carrier-phonon coupling the population ratio
${\overline{f}_b \over \overline{f}_a}$ decreases. Such a behavior
can be physically described in terms of a phonon-induced
renormalization of the interlevel energy splitting
$\Delta\epsilon$. Indeed, such renormalized transition energy can
be also obtained from the imaginary parts of the non-zero
eigenvalues of the effective Liouville operator in
(\ref{calL-4D}).

We finally stress that such bidimensional dressed basis as well as the corresponding energy shift can be
regarded as the simplest example of polaronic phase-coherence and state renormalization (see below).

\section{Polaronic coherence in biased semiconductor superlattices}\label{s-PC}

As a concrete example of scattering-induced single-particle phase
coherence in state-of-the-art semiconductor nanostructures, let us
consider high-field transport in biased semiconductor
superlattices. As extensively discussed in Ref.~\onlinecite{PRBC},
the treatment of carrier-phonon scattering in the presence of
strong electric fields requires a gauge-invariant formulation of
the problem; the density-matrix treatment proposed in
Ref.~\onlinecite{PRBC} has shown that the only single-particle
basis $\alpha$ in which the Markov limit is properly defined is
the Wannier-Stark one.

In order to gain more insight into the single-particle phase
coherence previously discussed, we have performed fully
three-dimensional calculations of high-field charge transport in
state-of-the-art semiconductor nanostructures. In particular, aim
of our description was to properly treat ---in addition to the
carrier quantum confinement along the growth direction--- the
in-plane energy-relaxation and thermalization dynamics.

As prototypical system, we consider a state-of-the-art GaAs-based
nanometric superlattice. More specifically, we have performed a
detailed investigation of the 30\AA\ GaAs/Al$_{0.3}$Ga$_{0.7}$As
superlattice structure shown in Fig.~\ref{f:sl1}. The
single-particle carrier states $\{\vert\alpha\rangle\}$ are
described within the usual envelope-function approximation in
terms of a space-independent effective mass $m^*$. They come out
to be products of two-dimensional plane waves and one-dimensional
envelope functions:
\begin{equation}\label{ef}
\phi_\alpha({\bf r}) = {1 \over \sqrt{\Omega}} e^{i{\bf k}_\perp
\cdot {\bf r}_\perp}
\phi^\parallel_{\alpha_\parallel}(r_\parallel) \ ,
\end{equation}
$\Omega$ denoting a suitable normalization area.
\begin{figure}
\unitlength1mm \ \par\noindent
\begin{picture}(80,93)
\put(5,-2){\psfig{figure=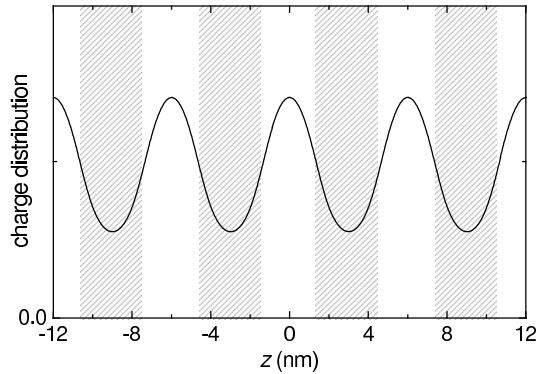,angle=0,width=70mm}}
\end{picture}
\caption{\label{f:sl1} Schematics of the prototypical 30\AA/30\AA\
GaAs/Al$_{0.3}$Ga$_{0.7}$As superlattice structure considered in
our simulated experiments: Real-space periodic nanostructure
profile (shaded regions correspond to barriers) and charge
distribution corresponding to the ground-state ($k_{\parallel}=0$)
envelope function in the field-free case[($n(r_{\parallel})
\propto \vert
\phi^0_{k_\parallel=0,\nu=1}(r_{\parallel})\vert^2$)]. }
\end{figure}

In the free-field case, the envelope functions $\phi^\parallel$ in
(\ref{ef}) reduce to one-dimensional Bloch states
$\phi^\parallel_{k_\parallel\nu}$ corresponding to the periodic
heterostructure potential reported in the inset of
Fig.~\ref{f:sl2}. As we can see, we deal with a relatively small
band-edge discontinuity ($V_\circ = 250$\,meV). The latter,
combined with a barrier width of 30\AA\ gives rise to significant
inter-well carrier tunnelling. This is confirmed by the field-free
ground-state charge distribution (solid curve in
Fig.~\ref{f:sl1}), which shows a clear fingerprint of carrier
delocalization.
\begin{figure}
%\unitlength1mm \ \par\noindent
%\begin{picture}(80,70)
%\put(-2,-12){\psfig{figure=fig3new.eps,angle=0,width=80mm}}
%\end{picture}
\includegraphics[width=0.45\textwidth]{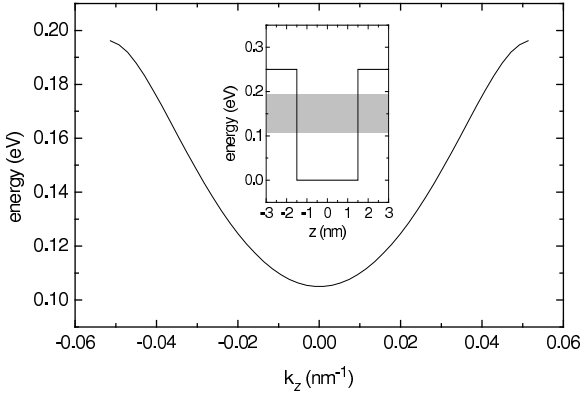}
\caption{\label{f:sl2} Single-miniband diagram (miniband width of
about 20 meV) corresponding to the superlattice structure depicted
in Fig.~\ref{f:sl1}. The superlattice potential profile (band-edge
discontinuity of 0.25 eV) is also shown in the inset. }
\end{figure}

The inter-well coupling displayed in Fig.~\ref{f:sl1} should
translate into a dispersive energy-momentum relation along the
growth direction. This is confirmed by the superlattice miniband
profile reported in Fig.~\ref{f:sl2}. As we can see, we deal with
a carrier miniband only: Its width amounts to about 90 meV and it
is therefore larger than the longitudinal optical (LO)-phonon
energy ($\simeq 36$ meV).

In the presence of an applied field $F$ along the growth
direction, the one-dimensional envelope functions $\phi^\parallel$
within the scalar-potential gauge  correspond to the usual
Wannier-Stark states.\cite{PRBC} The latter are displayed in
Fig.~\ref{f:sl3} for different values of $F$. As we can see, for
increasing values of the field we deal with an increasing state
localization and a corresponding suppression of inter-well
single-particle tunnelling.
\begin{figure}
%\unitlength1mm \ \par\noindent
%\begin{picture}(80,150)
%\put(-8,-13){\psfig{figure=fig4new.eps,angle=0,width=95mm}}
%\end{picture}
\includegraphics[width=0.5\textwidth]{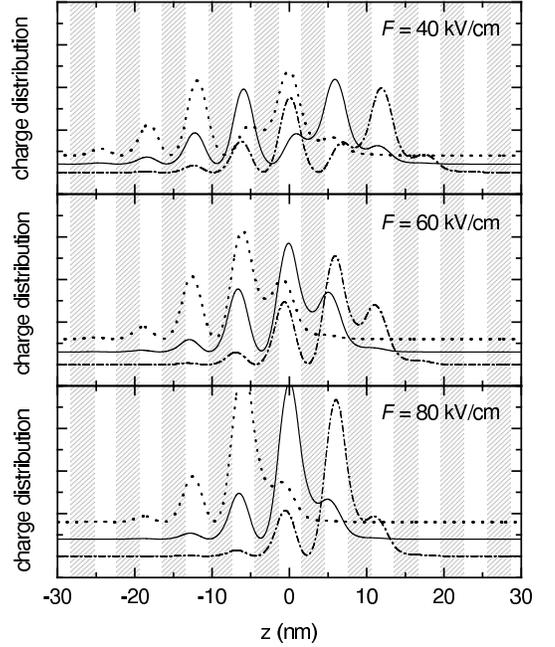}
\caption{\label{f:sl3} Charge distribution corresponding to the
Wannier-Stark states in the superlattice structure of
Fig.~\ref{f:sl1} for different values of the applied field $F$.
For each field three states are displayed: $n = -1$ (dotted
curve), $n=0$ (solid curve), and $n = 1$ (dash-dotted curve). }
\end{figure}

Primary goal of our simulated experiments was the study of the
current-voltage characteristics of the superlattice structure
previously introduced, in the presence of carrier-LO phonon
scattering. In particular, a thermal bath of three-dimensional
(bulk) dispersionless longitudinal polar optical phonons has been
assumed, employing the carrier-phonon interaction model given in
Ref.~\onlinecite{Marburg}. We have therefore evaluated the carrier
drift velocity as a function of the applied field. The latter can
be readily computed according to Eq.~(\ref{SP2}), using as single
particle quantity the velocity operator:
\begin{equation}\label{v}
{\bf A}^{sp} = {\bf v} = {{\bf P} \over m^*} = -{i\hbar \over m^*} \nabla \ .
\end{equation}
More specifically, we get:
\begin{equation}\label{SP3-bis}
v^d = \sum_{\alpha\alpha'} v_{\alpha\alpha'} \rho^{sp}_{\alpha'\alpha} \ ,
\end{equation}
where $\rho^{sp}_{\alpha\alpha'}$ is the single-particle density
matrix, and $v_{\alpha\alpha'}$ denote the matrix elements of the
velocity operator (\ref{v}) within our $\alpha$ representation.

\begin{figure}
\includegraphics[width=0.4\textwidth]{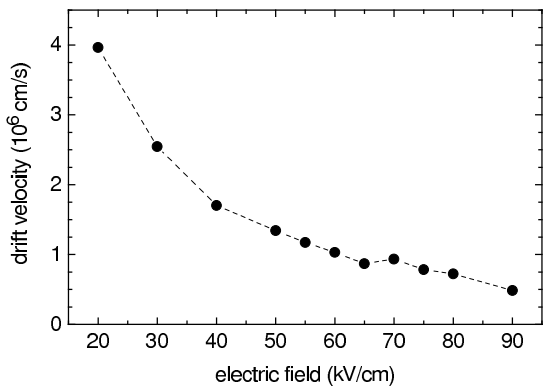}
\caption{\label{f:sl4} Steady-state carrier drift velocity as a
function of the applied field for the superlattice structure of
Fig.~\ref{f:sl1} at room temperature, in the low-density limit.
The line is a guide to the eye.}
\end{figure}
Figure \ref{f:sl4} shows the steady-state carrier drift velocity
as a function of the applied field for the superlattice structure
of Fig.~\ref{f:sl1} at room temperature in the low-density limit.
The peak at $\simeq 70$ kV/cm corresponds to the phonon resonance,
i.e., for this value of the applied field, the Wannier-Stark
energy $eFd$ ---$d$ being the superlattice period--- is equal to
the LO-phonon energy. We stress that such feature well agrees with
the results obtained with the non-equilibrium Green's function
treatment by Jauho and co-workers.\cite{WackerPRL} As we can see,
we get drift-velocity values significantly different from zero.
This remark is particularly important in view of the fact that
such field-induced current is entirely due to the presence of
non-diagonal density-matrix elements; indeed, within the
semiclassical limit previously introduced we have
$\rho^{sp}_{\alpha\alpha'} = f^{sp}_\alpha
\delta_{\alpha\alpha'}$, and therefore Eq. (\ref{SP3-bis}) reduces
to:
\begin{equation}\label{SP3-ter}
v^d = \sum_{\alpha} v_{\alpha\alpha} f^{sp}_\alpha \ .
\end{equation}
As we can see, the only matrix elements of the velocity operator
involved are the diagonal ones, i.e., $v_{\alpha = \alpha'}$; for
localized as well as periodic basis states $\alpha$ such diagonal
matrix elements are known to vanish, and therefore within the
semiclassical limit the drift velocity in (\ref{SP3-bis}) is equal
to zero, i.e., no current.

It is then easy to conclude that the significantly large values of
the drift velocity reported in Fig. \ref{f:sl4} are entirely due
to the non-diagonal density-matrix elements $\rho^{sp}_{\alpha \ne
\alpha'}$ entering Eq. (\ref{SP3-bis}); in turn such non-diagonal
contributions reflect a scattering induced phase coherence between
our Wannier-Stark states. As a result, we can say that the
significant carrier drift velocity in Fig. \ref{f:sl4} is an
unambiguous fingerprint of phonon-induced phase coherence in
semiconductor superlattices.

\section{Summary and conclusions}\label{s-SC}

We have presented a general density-matrix formulation of
quantum-transport phenomena in semiconductor nanostructures
typically employed in the design of state-of-the-art quantum
devices. More specifically, contrary to the conventional
single-particle correlation expansion, we have investigated
separately the effects of the Markov limit and of the reduction
procedure. Our fully operatorial approach has allowed us to better
identify the general properties of the scattering superoperators
entering our effective quantum-transport theory at various
description levels, e.g., $N$ electrons-plus-quasiparticles, $N$
electrons only, and single-particle picture.

In addition to coherent transport phenomena characterizing the
transient response of the system, the proposed theoretical
description has allowed the study of scattering induced phase
coherence in steady-state conditions. In particular, based on the
proposed approach we have investigated
polaronic effects in
strongly biased semiconductor superlattices.

One important conclusion of our theoretical analysis is that the
Markov limit alone ---i.e., not combined with the diagonal or
semiclassical approximation--- may lead to totally non-physical
results: (i) within the global (carrier + quasiparticle) picture
the Markov limit introduces a fictitious energy
redistribution/relaxation and dephasing which has no physical
counterpart within a closed-system scenario; (ii) both for the
global and for the reduced pictures, the resulting effective
Liouville superoperators do not correspond, in general, to a CP
map, thus preventing from any ``robust'' time-dependent solution
of our quantum-transport equation.

A second important remark is that the combination of the Markov
limit with the semiclassical approximation
---i.e., the neglect of non-diagonal density-matrix elements---
leads, also for the global description, to Boltzmann-like
equations whose solution can be shown to be positive-definite at
any time.

In spite of the fact that our density-matrix formulation within
the usual Markov limit does not translate into CP maps, our
analysis has shown that within the small-coupling limit the
proposed quantum-transport equations are always characterized by a
positive-definite steady-state solution.

\section*{Acknowledgments}

We are grateful to Massimo Fischetti, Carlo Jacoboni, Tilmann
Kuhn, and Paolo Zanardi for stimulating and fruitful discussions.

\appendix

\section{General properties of the scattering superoperator}
\label{a-PDC}

Aim of this appendix is to discuss the general structure and
symmetry properties which characterize the global (carrier +
quasiparticle) in- and out-scattering superoperators in
(\ref{in2})-(\ref{out2}), their reduced versions in
(\ref{in2_c})-(\ref{out2_c}), and the single-particle ones in
(\ref{LDL-in})-(\ref{LDL-out}).

To this end, let us consider first the semiclassical limit
introduced in Sect.~\ref{ss-SL}. For the global picture %[see Eq.~(\ref{SL3})]
as well as for the reduced and single-particle descriptions
%[see Eqs.~(\ref{SL5}) and (\ref{LDL-SL})]
we deal with a Boltzmann transport equation of the form:
\begin{equation}\label{BTE1}
{d f_i \over dt} =
\sum_{i'} \left(
P_{ii'} f_{i'} - P_{i'i} f_i
\right)\ .
\end{equation}
In analogy with the definition of the scattering superoperator $\Gamma$
in (\ref{Gamma-lambda1}), the transport equation (\ref{BTE1}) can be rewritten as:
\begin{equation}\label{BTE2}
{d f_i \over dt} =
\sum_{i'}
\gamma_{ii'} f_{i'} \ ,
\end{equation}
where
\begin{equation}\label{gamma}
\gamma_{ii'} = \gamma^{\rm in}_{ii'} - \gamma^{\rm out}_{ii'} =
P_{ii'} -
\delta_{ii'} \sum_{i''} P_{i''i}
\end{equation}
is again the difference of in- and out-contributions.
As a first remark, we notice that (i) the in-contribution coincides with the semiclassical scattering
rate $P_{ii'}$ and (ii) the out-contribution is a function of the in-one:
\begin{equation}\label{out-vs-in}
\gamma^{\rm out}_{ii'} = \delta_{ii'} \sum_{i''} \gamma^{\rm in}_{i''i} \ .
\end{equation}

The Boltzmann transport equation in (\ref{BTE1}) is supposed to preserve the total number of particles
$N = \sum_i f_i$, i.e.,
\begin{equation}\label{N}
{d N \over dt} = \sum_i {d f_i \over dt} =
\sum_{ii'}
\gamma_{ii'} f_{i'} = 0 \ .
\end{equation}
The above conservation law ---which corresponds to the trace-preserving property of the quantum-mechanical
case--- needs to be verified for any carrier distribution $f_i$;
this requires that:
\begin{equation}\label{TP1}
\sum_i \gamma_{ii'} = 0 \ ,
\end{equation}
which can be readily verified starting from the definition of the operator $\gamma$ in (\ref{gamma}).

Let us now consider the eigenvalue problem for the semiclassical scattering operator $\gamma$:
\begin{equation}\label{gamma-eigen}
\sum_{i'} \gamma_{ii'} f_{i'} = \Lambda f_i \ .
\end{equation}
As a first property, the eigenvalue spectrum is non-positive
definite: $\Lambda \leq 0$; secondly, the particle-preserving
property in (\ref{TP1}) requires the existence of the $\Lambda =
0$ eigenvalue, whose eigenvector $f_i$ corresponds to the
steady-state solution of the problem.

We shall finally discuss the positive-definite character of the semiclassical carrier dynamics. Starting
from the Boltzmann equation in (\ref{BTE2}), we can express the carrier distribution at time $t$ in terms
of the semiclassical propagator $s_{ij}(t-t_0)$:
\begin{equation}\label{sp1}
f_i(t) = \sum_{j} s_{ij}(t-t_0) f_{j}(t_0) \ ,
\end{equation}
where $s_{ij}(0) = \delta_{ij}$ and
\begin{equation}\label{sp2}
{d s_{ij} \over dt} =
\sum_{i'}
\gamma_{ii'} s_{i'j} \ .
\end{equation}
By inserting into the above equation the explicit form of the scattering operator $\gamma$ in (\ref{gamma})
we get:
\begin{equation}\label{sp3}
{d s_{ij} \over dt} =
-\overline{\gamma}_i s_{ij} +
\sum_{i'}
P_{ii'} s_{i'j} \ ,
\end{equation}
where $\overline{\gamma}_i = \sum_{i'} P_{i'i}$ denotes the total or out-scattering rate for state $i$.
The above differential equation can be formally integrated according to:
\begin{equation}\label{sp4}
s_{ij}(t) = e^{-\overline{\gamma}_i (t-t_0)} \delta_{ij} + \int_{t_0}^t dt' e^{-\overline{\gamma}_i (t-t')}
\sum_{i'} P_{ii'} s_{i'j}(t') \ .
\end{equation}
This formal solution ---also known as Cham\-ber's
for\-mu\-la\-tion~\cite{RPJ}--- is the starting point of
the semiclassical Neumann expansion. Indeed by iteratively
substituting Eq.~(\ref{sp4}) into itself, we obtain:
\begin{widetext}
\begin{equation}\label{sp5}
s_{ij}(t) = \sum_{n=0}^\infty
\int_{t_0}^t dt_1 \dots \int_{t_0}^{t_{n-1}} dt_n \sum_{i_1, \dots,i_{n-1}}
e^{-\overline{\gamma}_i(t-t_1)} P_{ii_1} e^{-\overline{\gamma}_{i_1}(t_1-t_2)} \dots P_{i_{n-1}j}
e^{-\overline{\gamma}_j (t_n-t_0)} \ .
\end{equation}
\end{widetext}
The above expansion can be regarded as the semiclassical
counterpart of the Neumann series in
(\ref{Neumann}).\cite{RPJ} By recalling that the
scattering rates $P_{ii'}$ and $\overline{\gamma}_i$ are
positive-definite quantities, it is easy to recognize that all the
terms of the above Neumann expansion are non-negative (i.e.,
greater or equal to zero); therefore, we can conclude that all the
matrix elements $s_{ij}(t)$ of the semiclassical propagator are
positive-definite, which in turn ensures that the distribution
function $f_i$ in (\ref{sp1}) is itself positive-definite. As
anticipated, we see that the combination of the Markov limit with
the semiclassical or diagonal approximation leads to a completely
positive time evolution, both for the global (carrier +
quasiparticle) case and for the reduced as well as single-particle
pictures.
\smallskip\par

Let us now move to the quantum-mechanical case. A closer
inspection of the in- and out-scattering superoperators in
(\ref{in2})-(\ref{out2}) as well as of their reduced versions in
(\ref{in2_c})-(\ref{out2_c}) and (\ref{LDL-in})-(\ref{LDL-out})
reveals that they have a general structure of the form:
\begin{equation}\label{in-str}
\Gamma^{\rm in}_{i_1i_2,i'_1i'_2} =
{\cal P}^{\rm in}_{i_1i_2,i'_1i'_2} +
{\cal P}^{\rm in *}_{i_2i_1,i'_2i'_1} \ ,
\end{equation}
\begin{equation}\label{out-str}
\Gamma^{\rm out}_{i_1i_2,i'_1i'_2} =
{\cal P}^{\rm out}_{i_1i_2,i'_1i'_2} +
{\cal P}^{\rm out *}_{i_2i_1,i'_2i'_1} \ ,
\end{equation}
with
\begin{equation}\label{calP-in}
{\cal P}^{\rm in}_{i_1i_2,i'_1i'_2} =
{2\pi \over \hbar^2}
g_{i_1i'_1} g^*_{i_2i'_2} {\cal D}^*_{i_2i'_2}
\end{equation}
and
\begin{equation}\label{calP-out}
{\cal P}^{\rm out}_{i_1i_2,i'_1i'_2} =
\delta_{i_2i'_2} \sum_{i''} {\cal P}^{\rm in *}_{i''i'',i_1i'_1} \ .
\end{equation}
Exactly as in the semiclassical theory previously recalled [see Eq.~(\ref{out-vs-in})], the out-scattering
superoperator is fully determined by the in-one. The latter, in turn, can be regarded as the quantum-mechanical
(i.e., nondiagonal) generalization of the conventional Fermi's golden rule; indeed, its semiclassical
contributions ($i_1i'_1 = i_2i'_2$) are of the form:
\begin{equation}\label{SC}
\Gamma^{\rm in}_{ii,i'i'} =
{2\pi \over \hbar^2}
\left|g_{ii'}\right|^2
\left({\cal D}_{ii'} + {\cal D}_{i'i}\right) \ .
\end{equation}

Similar to the particle-conservation law in (\ref{N}), in the quantum-mechanical case we ask the
scattering superoperator
\begin{equation}\label{in-out}
\Gamma_{i_1i_2,i'_1i'_2} =
\Gamma^{\rm in}_{i_1i_2,i'_1i'_2} -
\Gamma^{\rm out}_{i_1i_2,i'_1i'_2}
\end{equation}
to be trace preserving, i.e.,
\begin{equation}\label{TP2}
{d \over dt} \rmtr\left\{\hat{\rho}\right\} =
\rmtr\left\{\Gamma\left(\hat{\rho}\right)\right\} = 0 \ .
\end{equation}
Rewriting the above equation within our generic $i$-basis we get:
\begin{equation}\label{TP3}
\sum_i \sum_{i'_1i'_2} \Gamma_{ii,i'_1i'_2} \rho_{i'_1i'_2} = 0 \ .
\end{equation}
This relation should hold for any generic $\rho$; this requires that:
\begin{equation}\label{TP4}
\sum_i \Gamma_{ii,i'_1i'_2} = 0 \ .
\end{equation}
The above trace-preserving condition can be readily verified starting from the general structure
of the in- and out-scattering superoperators in (\ref{in-str})-(\ref{out-str}). We stress that
such property is verified for any form of the in-scattering superoperator ${\cal P}^{\rm in}$,
i.e., does not depend on the specific structure of the scattering operator in (\ref{calP-in}).

Let us now come to the eigenvalue problem for the scattering superoperator $\Gamma$ and for the
corresponding effective Liouville superoperator ${\cal L}$ [see Eqs.~(\ref{LDL-final}) and (\ref{calL_sp})].
By denoting with $j = i_1i_2$ the generic pair of indices for our density matrix, we have:
\begin{equation}\label{eigen-bis}
{\cal L}_{jj'} \rho_{j'} = \Lambda \rho_j \ .
\end{equation}
Contrary to the semiclassical picture,
now the eigenvalues $\Lambda$ are complex quantities and, more important their real parts may
assume positive as well as negative values. This is a clear indication of the fact that we are
not dealing with a CP map (see below).
However, the trace-preserving property in (\ref{TP4}) requires also for the quantum-mechanical
case the existence of the $\Lambda = 0$ eigenvalue, whose eigenvector $\rho_j = \rho_{i_1i_2}$
corresponds to the steady-state solution of the problem.

We shall finally discuss the possible positive-definite character of our density-matrix $\rho$.
In very general terms we can express the density matrix at time $t$ via a quantum-mechanical
propagator or quantum-mechanical map
${\cal S}$:
\begin{equation}\label{qp1}
\rho_{i_1i_2}(t) = \sum_{j_1j_2} {\cal S}_{i_1i_2,j_1j_2}(t-t_0) \rho_{j_1j_2}(t_0) \ ,
\end{equation}
where ${\cal S}_{i_1i_2,j_1j_2}(0) = \delta_{i_1i_2,j_1j_2}$ and
\begin{equation}\label{qp2}
{d \over dt} {\cal S}_{i_1i_2,j_1j_2} = \sum_{i'_1i'_2} {\cal L}_{i_1i_2,i'_1i'_2}
{\cal S}_{i'_1i'_2,j_1j_2} \ .
\end{equation}
Using the more compact operatorial notation we have:
\begin{equation}\label{qp3}
\hat\rho(t) = {\cal S}\left(\hat \rho(t_0)\right)
\end{equation}
with
\begin{equation}\label{qp4}
{d {\cal S} \over dt} = {\cal L} {\cal S} \ .
\end{equation}
It is well known that the most general form of a CP map is given by:
\begin{equation}\label{qp5}
{\cal S}\left(\hat\rho(t_0)\right) = \sum_k \hat{\cal M}_k(t-t_0) \hat\rho(t_0) \hat{\cal M}^\dagger(t-t_0) \ ,
\end{equation}
where the generic set of operators $\hat{\cal M}_k$ ---usually
referred to as ``Krauss operators''--- should obey the
normalization condition
\begin{equation}\label{Krauss}
\sum_k \hat{\cal M}_k \hat{\cal M}^\dagger_k = \hat{\cal I} \ .
\end{equation}

A particular class of CP maps is given by the so-called Lindblad
maps, whose effective Liouville superoperators are of the form:
\begin{equation}\label{qp6}
{\cal L}\left(\hat\rho\right) = {1 \over i\hbar} [\hat H_\circ,\hat\rho] + \sum_k
\left(
\left[\hat{\cal A}_k \hat\rho, \hat{\cal A}^\dagger_k\right]
+
\left[\hat{\cal A}_k,\hat\rho \hat{\cal A}^\dagger_k\right]
\right) \ .
\end{equation}
For the particular case of Hermitian operators $\hat{\cal A}_k$, the Lindblad form in Eq.~(\ref{LDC})
is recovered.

Let us now try to compare the effective scattering dynamics described by the transport equation in
(\ref{LDL-final}) with the general Krauss expansion in (\ref{qp5}) and with the Lindblad form in (\ref{qp6}).

For the global description, the double-commutator in
(\ref{Ltilde}) is not Lindblad-like; indeed, as already pointed
out, its eigenvalue spectrum is in general not negative-definite.
This tells us that, while the steady-state solution of the
Lindblad form in (\ref{LDC}) is again proportional to the identity
operator $\hat {\cal I}$, its dynamical
evolution in general does not preserve the positive-definite
character of $\hat\rho$ and may also exhibit singularities. Such
anomalous and totally non-physical behavior
---not present within the semiclassical or Boltzmann picture--- is
again the result of the Markov limit.

Also for the case of the reduced and single-particle descriptions, their effective Liouville operators
derived in Sects.~\ref{ss-RD} and \ref{ss-SPD} do not correspond, in general, to CP maps, and its time
evolution does not preserve the positive-definite character of $\hat\rho$.

In spite of the fact that, in general, we are not dealing with CP
maps, it is possible to show that for all the transport equations
derived in the paper we have a steady-state solution. Moreover, in
the so-called small-coupling limit, i.e., when the perturbation
Hamiltonian $\hat H'$ is small compared to the noninteracting
Hamiltonian $\hat H_\circ$, it is possible to show that such
steady-state ($\Lambda = 0$) solution is always positive-definite
[see Eq.~(\ref{PD})].

% \begin{acknowledgments}

% \end{acknowledgments}

\newpage


\begin{thebibliography}{xx}
\bibitem{note-entropy}
The quantum entropy in (\ref{entropy}) is time-independent since
the trace is invariant under the unitary transformation
(\ref{rho_i}).
\end{thebibliography}

{}

\newpage

\end{document}